\documentclass[11pt,journal,draftcls,twocolumn,peerreviewca]{IEEEtran}

\usepackage{amsfonts}
\usepackage{amsmath}
\usepackage{amssymb}
\usepackage{graphicx}
\usepackage{color, soul}
\usepackage{stfloats}
\usepackage{setspace}
\usepackage{algorithm}
\usepackage{algorithmic}
\usepackage{float}

\usepackage[amsmath,thmmarks]{ntheorem}
\usepackage{theorem}
\usepackage[justification=centering]{caption}

\newtheorem{proposition}{Proposition}

\theoremheaderfont{\sc}\theorembodyfont{\upshape}
\theoremstyle{nonumberplain}
\theoremseparator{}
\theoremsymbol{\rule{1ex}{1ex}}

\begin{document}

\title{Signal Amplitude Estimation and Detection from Unlabeled Binary Quantized Samples}

\author{Guanyu Wang, Jiang Zhu, Rick S. Blum, Peter Willett,\\ Stefano Marano, Vincenzo Matta, Paolo Braca}

\date{}
\maketitle
\begin{abstract}
Signal amplitude estimation and detection from unlabeled quantized binary samples are studied, assuming that the order of the time indexes is completely unknown. First, maximum likelihood (ML) estimators are utilized to estimate both the permutation matrix and unknown signal amplitude under arbitrary, but known signal shape and quantizer thresholds. Sufficient conditions are provided under which an ML estimator can be found in polynomial time and an alternating maximization algorithm is proposed to solve the general problem via good initial estimates. In addition, the statistical identifiability of the model is studied.

Furthermore, the generalized likelihood ratio test (GLRT) detector is adopted to detect the presence of signal. In addition, an accurate approximation to the probability of successful permutation matrix recovery is derived, and explicit expressions are provided to reveal the relationship between the number of signal samples and the number of quantizers. Finally, numerical simulations are performed to verify the theoretical results.

\end{abstract}
\section{Introduction}

In many systems, the data is transmitted with time information, which may sometimes be imprecise \cite{GPS, Humphreys1, Zeng1, Zhang2, Challa1, Schenato1, Millefiori}.
One example is the global positioning system (GPS) spoofing attack which can alter the time stamps on electric grid measurements \cite{GPS} to make them useless so that the data must be processed without time stamps. Since the exact form of civilian GPS signals is publicly known and the elements needed are inexpensive, building a circuit to generate signals to spoof the GPS is easy.
In \cite{Humphreys1}, a refined assessment of the spoofing threat is provided. In addition, the detailed information of receiver-spoofer architecture, its implementation and performance, and spoofing countermeasures are introduced. As a case study in \cite{Zeng1}, the impact of the GPS spoofing attack on wireless communication networks, more specifically, the frequency hopping code division multiple access (FH-CDMA) based ad hoc network, is investigated. A timing synchronization attack (TSA) is coined to the wide area monitoring systems (WAMSs), and its effectiveness is demonstrated for three applications of a phasor measurement unit (PMU) \cite{Zhang2}. In \cite{Challa1}, the out-of-sequence measurement (OOSM) problem where sensors produce observations that are sent to a fusion center over communication networks with random delays are studied, and a Bayesian solution is provided. The problem of random delay and packet loss in networked control systems (NCS) is studied in \cite{Schenato1}. In addition, a minimum error covariance estimator for the system is derived and two alternative estimator architectures are presented for efficient computation. In \cite{Millefiori}, the effect of an unknown timestamp delay in Automatic Identification System (AIS) is studied, and a method based on adaptive filtering is proposed.

In the above examples, the relative order of the data is unknown, i.e., the samples are unlabeled. Estimation and detection from unlabeled samples have drawn a great deal of attention recently \cite{Emiya1, Unnikrishnan1c, Pananjady1c, Pananjady2, Abid1, Haghighatshoar1c, Keller1, Marano1, Braca, Jiang4, Unnikrishnan1, Pananjady1, Haghighatshoar1}. In \cite{Emiya1}, it is shown that the convex relaxation based on a Birkhoff polytope approach does not recover the permutation matrix, and a global branch and bound algorithm is proposed to estimate the permutation matrix. In the noiseless case with a random linear sensing matrix, it is shown that the permutation matrix can be recovered correctly with probability $1$, given that the number of measurements is twice the number of unknowns \cite{Unnikrishnan1c, Unnikrishnan1}. In \cite{Pananjady1c, Pananjady1}, the noise is taken into account and a condition under which the permutation matrix can be recovered with high probability is provided. In addition, a polynomial time algorithm is proposed for a scalar parameter case.  Denoising linear regression model with shuffled data and additive Gaussian noise are studied in \cite{Pananjady2}, and the characterization of minimax error rate is provided. In addition, an algorithm for the noiseless problem is also proposed, and its performance is demonstrated on an image point-cloud matching task \cite{Pananjady2}. In \cite{Abid1}, several estimators are compared in recovering the weights of a noisy linear model from shuffled labels, and an estimator based on the self-moments of the input features and labels is introduced.
For unlabeled ordered sampling problems where the relative order of observations is known, an alternating maximization algorithm combined with dynamic programming is proposed \cite{Haghighatshoar1c}. In \cite{Marano1}, a signal detection problem where the known signal is permuted in an unknown way is studied.

Compared to the location parameter estimation problem ($x_i=\theta+w_i$) in \cite{Jiang4}, the model in this paper is a scale parameter estimation problem ($x_i=h_i\theta+w_i$), in which $h_i,~i=1,\cdots,K$ is the shape of a signal, and $\theta$ is an amplitude of signal. As a result, the scale parameter estimation problem is much more difficult than the location estimation in several aspects, and the scale parameters are especially relevant in relation to the mislabeling/permutation issue. First, the model in \cite{Jiang4} is always identifiable, while our model may be unidentifiable, as shown later. Second, the problem in \cite{Jiang4} can be solved efficiently via simple sorting, while we can only prove that problem in this paper can be solved efficiently provided certain conditions are satisfied. Third, good initial points are proposed to improve the performance of alternating maximization algorithm. Furthermore, we provide an approximation to the probability of successful permutation matrix recovery, which reveals the relationship between the length of signal and the number of quantizers.

In this paper, we focus on the problems of scale estimation and signal detection from unlabeled quantized samples. The main contribution of this work can be summarized as follows. First, a sufficient condition for the existence of a polynomial time algorithm is provided for the unlabeled estimation problem, and the model is shown to be unidentifiable in some special cases. Second, good initial points are provided to improve the performance of an alternating maximization algorithm. And third, we provide analytic approximations on probability of permutation matrix recovery in the case of known signal amplitude, which can be used to predict when the permutation matrix can be correctly recovered.



The organization of this paper is as follows. In Section \ref{setup}, the problem is described. Background on ML estimation and generalized likelihood ratio test (GLRT) detection from labeled data are presented in Section \ref{pre}. In section \ref{section_est}, the model identifiability is studied, and the estimation problem from unlabeled data is studied. Section \ref{det} extends the detection work to unlabeled data, and derives an approximate analytic formula for permutation matrix recovery probability. Finally, numerical results are presented in Section \ref{simulation}, and conclusion follows in Section \ref{con}.

Notation: The $K\times 1$ vector of ones is ${\mathbf 1}_K$.  For an unknown deterministic parameter $\theta$, $\theta_0$ denotes its true value. For an unknown permutation matrix $\boldsymbol\Pi$, $\boldsymbol\Pi_0$ denotes its true value. For a random vector $\mathbf y$, ${ p}({\mathbf y};\theta)$ denotes the probability density function (PDF) of $\mathbf y$ parameterized by $\theta$, and ${\rm E}_{\mathbf y}[\cdot]$ denotes the expectation taken with respect to $\mathbf y$. Let ${\mathcal N}(\mu,\sigma^2)$ denote a Gaussian distribution with mean $\mu$ and variance $\sigma^2$. Let $\Phi(\cdot)$ and $\varphi(\cdot)$ denote the cumulative distribution function (CDF) and probability density function (PDF) of a standard Gaussian random variable respectively. Let ${\mathcal U}(a,b)$ denote an uniform distribution, whose minimum and maximum values are $a$ and $b$. Let ${\mathcal B}(N,p)$ denote a binomial distribution, where $N$ and $p$ denote the number of trials and the probability of event, respectively.

\section{Problem Setup}\label{setup}

\begin{figure}[h!t]
\centering
\includegraphics[width=2.6in]{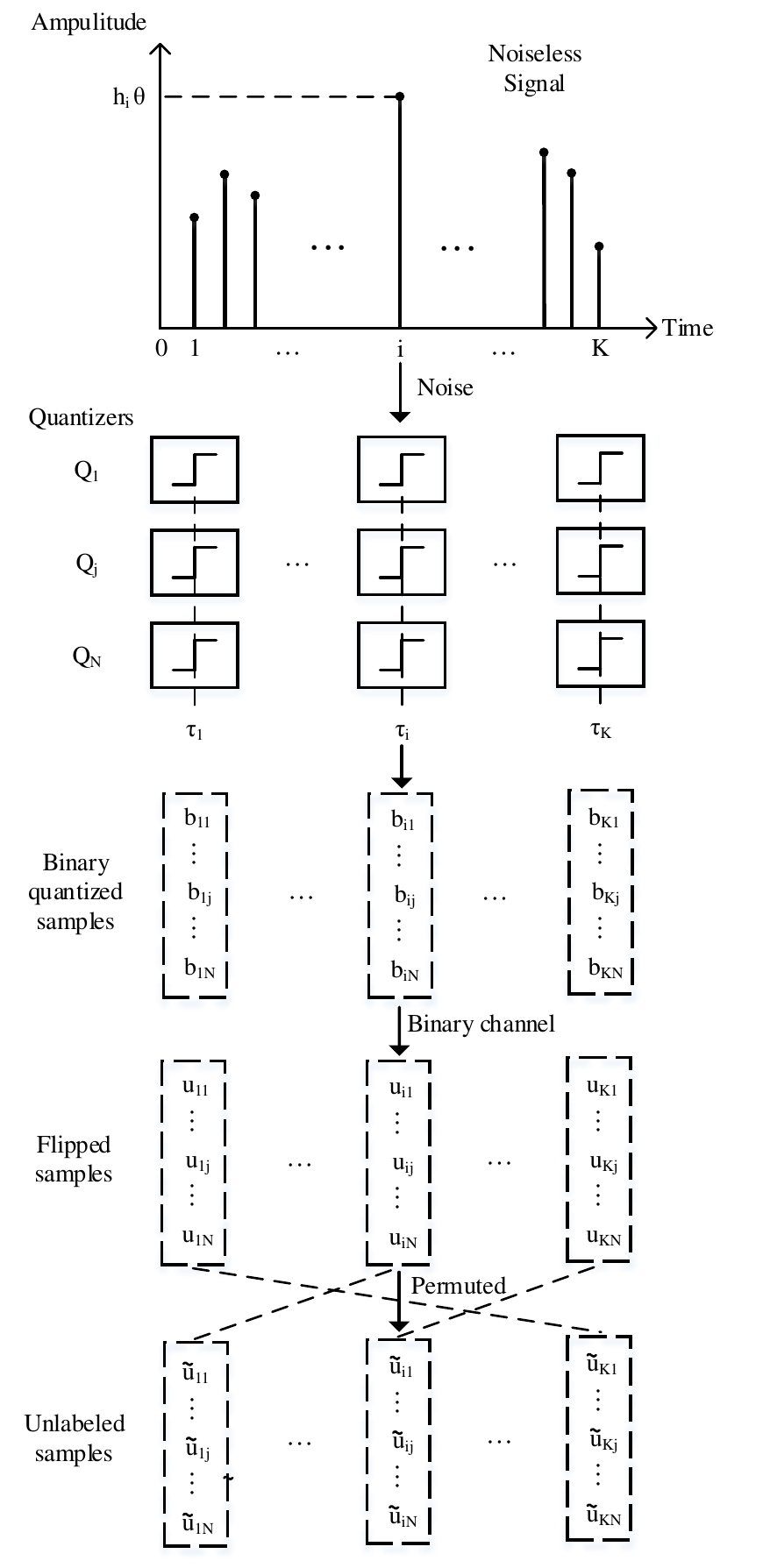}
\caption{System diagram of unlabeled binary quantized samples generation.}
\label{Model}
\end{figure}

Consider a signal amplitude estimation and detection problem where a collection of $N$ binary quantizers generate binary quantized samples which will be utilized to estimate the unknown scaling factor $\theta$ of a $K$ length signal and detect the presence of the signal, as shown in Fig. \ref{Model}.
The binary quantized samples $b_{ij}$ are obtained via
\begin{equation}\label{generation_sample}
\begin{aligned}
b_{ij}=Q_i(h_i{\theta}+w_{ij}), \quad i=1,\cdots,K, \quad j=1,\cdots,N,
\end{aligned}
\end{equation}
and the corresponding hypothesis problem can be formulated as
\begin{align}
\begin{cases}
&\!\!\!{\mathcal H}_0:b_{ij}=Q_i(w_{ij}),  \; i=1,\cdots,K, \; j=1,\cdots,N,\notag\\
&\!\!\!{\mathcal H}_1:b_{ij}=Q_i(h_i{\theta}+w_{ij}),\; i=1,\cdots,K, \; j=1,\cdots,N,\notag
\end{cases}
\end{align}
where $i$ and $j$ respectively denote one of the $K$ time indexes and one of the $N$ quantizers, $h_i$ is the known coefficient characterizing the signal shape, $w_{ij}$ is the i.i.d. noise drawn from the $\sigma_w^2$-variance distribution whose PDF is $f_w(x/\sigma_w)/\sigma_w$ and CDF is $F_w(x/\sigma_w)$, where $f_w(x)$ and $F_w(x)$ are the corresponding unit-variance PDF and CDF, and $Q_i(\cdot)$ implies a binary quantizer which produces $1$ if the argument is larger than a scalar threshold $\tau_i$ and $0$ otherwise. The thresholds of $N$ quantizers are identical given any time index\footnote{Here we have thresholds fixed across quantizers and varying with time, with permutation across time. We could, equivalently, have fixed thresholds of quantizers across time but varying across sensors and permuted across sensors. Mathematically, it is the same problem and the formulation could as easily encompass it.}. We assume that the PDF $f_w(w)$ is log-concave, which is often met in practice such as Gaussian distributions, and the thresholds of $N$ quantizers are identical given any time index. We assume that the PDF $f_w(w)$ is logconcave, which is often met in practice by, for example, the Gaussian distribution.

The quantized data $\{ b_{ij} \}$ are transmitted over a binary channel with flipping probabilities $q_0$ and $q_1$ which are defined as ${\rm Pr}(u_{ij}=1|b_{ij}=0)=q_0$ and ${\rm Pr}(u_{ij}=0|b_{ij}=1)=q_1$, where $u_{ij}$ is the sample received at the output of the channel, which we call the fusion center (FC) \cite{Ozdemir1}.

We assume that all the sets of data $\{{u}_{ij}\}_{j=1}^N$ are transmitted to the FC with permuted time indexes.
Accordingly, the FC receives sets of data, say $\{\tilde u_{ij}\}_{j=1}^N$, in which the time reference (represented by the index $i$) is invalid. Specifically, the FC does not know which time index the data $\{\tilde{u}_{ij}\}_{j=1}^N$ belongs to, but knows that $\{\tilde{u}_{ij}\}_{j=1}^N$ belongs to one of the $K$ time indexes. Let us introduce the matrix $\mathbf U$ whose $(i,j)-$th entry is $u_{ij}$. Then, the unlabeled samples can be collected in a matrix $\widetilde{\mathbf U}$, as follows:
\begin{align}\label{unlabel}
\widetilde{\mathbf U}={\boldsymbol \Pi}{\mathbf U},
\end{align}
where ${\boldsymbol \Pi}\in{\mathbb R}^{K\times K}$ is an unknown permutation matrix; that is, a matrix of \{0,1\} entries in which each row and each column
sums to unity. We assume that $\theta$ is constrained to an interval $[-\Delta,\Delta]$, for algorithm and theoretical reasons \cite{Papa_tit}.
\section{Preliminaries}\label{pre}
In this section, standard materials of parameter estimation and signal detection using labeled data are presented.
\subsection{Maximum Likelihood Estimation}\label{mle_pre}
The probability mass function (PMF) of $u_{ij}$ can be calculated as

\begin{equation}\label{y_probability}
\begin{aligned}
{\rm Pr}(u_{ij}=1)&=q_0+(1-q_0-q_1)F_w\left(\frac{h_i{\theta}-\tau_i}{\sigma_w}\right)\triangleq p_i,\\
{\rm Pr}(u_{ij}=0)&=1-p_i.\\
\end{aligned}
\end{equation}
The PMF of $\mathbf U$ is
\begin{align}\label{LK}
p(\mathbf U;{\theta})=\prod_{i=1}^{K} \prod_{j=1}^{N} {\rm Pr}(u_{ij}=1)^{u_{ij}} {\rm Pr}(u_{ij}=0)^{(1-u_{ij})}.
\end{align}
Let $\eta_i$ denote the fraction of $u_{ij}=1$ in $\{u_{ij}\}_{j=1}^{N}$, i.e.,
\begin{align}\label{eta_def}
\eta_i=\sum_{j=1}^{N}u_{ij}/N.
\end{align}
Consequently, the log-likelihood function $l({\boldsymbol \eta};{\theta})$ is
\begin{align}\label{log_likelihood_function}
l(\boldsymbol \eta;{\theta})=N\sum_{i=1}^{K} (\eta_i \log p_i+(1-\eta_i)\log(1-p_i)),
\end{align}
where $p_i$ is given in (\ref{y_probability}).
Note that in an error free binary symmetric channel scenario, i.e., $q_0=q_1=0$ or $q_0=q_1=1$, the CDF $F_w(x)$ is log-concave as it is the integral of a log-concave PDF $f_w(x)$. Therefore maximizing the log-likelihood function is a convex optimization problem, which can be solved efficiently via numerical algorithms \cite{Ribeiro1, Ribeiro2, boyd}. For $0<q_0+q_1<2$, it is difficult to determine the convexity of the negative log-likelihood function. In this case all that can be guaranteed is a local optimum. As we show in numerical experiments, we found that the ML estimator using gradient descent algorithm works well and approaches the Cram\'{e}r Rao lower bound (CRLB).

In addition, the Fisher Information (FI) $I(\theta)$ is the expectation of the negative second derivative of the log-likelihood function $l({\boldsymbol \eta};{\theta})$ (\ref{log_likelihood_function}) taken with respect to $\theta$, i.e., \cite{Kassambook},
\begin{align}
I({\theta})\!&=\!-\frac{N(\!1\!-\!q_0\!-\!q_1\!)}{\sigma_w}\!\sum_{i=1}^{K}\!h_i\Bigg\{\!f_w\!\!\left(\!\frac{h_i\theta\!-\!\tau_i}{\sigma_w}\!\right){\rm E}_{{\boldsymbol \eta}}\!\!\left[\frac{\partial}{\partial \theta}\!\! \left(\frac{\eta_i}{p_i}\!-\!\frac{1\!-\!\eta_i}{1\!-\!p_i}\right)\!\right]\!\!+\!\!\frac{\partial}{\partial \theta}f_w\!\!\left(\!\frac{h_i\theta\!-\!\tau_i}{\sigma_w}\right) {\rm E}_{\boldsymbol \eta}\!\!\left[\frac{\eta_i}{p_i}\!-\!\frac{1\!-\!\eta_i}{1\!-\!p_i}\!\right]\!\!\Bigg\}\!\notag\\
&=\!\frac{N(\!1\!-\!q_0\!-\!q_1\!)^2}{\sigma_w^2}\!\sum_{i=1}^{K}\!\frac{ h_i^2 f_w^2\left(\frac{h_i{\theta}-\tau_i}{\sigma_w}\right)} {p_i(1-p_i)},\label{FI}
\end{align}
where (\ref{FI}) follows due to ${\rm E}_{\boldsymbol \eta}[\eta_i/p_i-(1-\eta_i)/(1-p_i)]=0$. Consequently, the CRLB is
\begin{align}\label{CRLBdef}
{\rm CRLB}(\theta)=1/I({\theta}),
\end{align}
which is later used as a benchmark performance for ML estimation from labeled data in Section \ref{simulation}.
\subsection{GLRT detection}
In the case of known $\theta$, the optimal detector according to the NP criterion is the log-likelihood ratio test \cite{Kay1}. For unknown $\theta$, the GLRT is usually adopted. Although there is no optimality associated with the GLRT, it appears to work well in many scenarios of practical interest \cite{Kay2}. The GLRT replaces the unknown parameter by its MLE and decides ${\mathcal H}_1$ if
\begin{equation}\label{GLRT_Detector}
\begin{aligned}
T_1({\boldsymbol\eta})={\underset{\theta\in[-\Delta,\Delta]} {\operatorname{max}}~ l({{\boldsymbol \eta}};{\theta})}
-{ l({{\boldsymbol \eta}};0)}>\gamma,
\end{aligned}
\end{equation}
where $\gamma$ is a threshold determined by the given false alarm probability $P_{FA}$.

\section{Estimation from unlabeled data}\label{section_est}
In this section, we study the estimation problem from unlabeled data. First, we delineate the model. The statistical identifiability is investigated, and the estimation problem is studied separately in the cases of known and unknown $\theta$.
\subsection{Maximum likelihood estimation}\label{MLE}
Introduce the function $\pi(\cdot)$ such that $m = \pi(i)$ if the permutation matrix ${\boldsymbol \Pi}$ in (\ref{unlabel}) maps the $i$th row of ${\mathbf U}$ to the $m$th row of $\tilde{\mathbf U}$. The PMF of $\tilde{\mathbf U}$ is
\begin{equation}
\begin{split}
&p(\widetilde{\mathbf U};\theta, {\boldsymbol \Pi})=\prod_{m=1}^{K} \prod_{j=1}^{N} {\rm Pr}(\tilde{u}_{mj}=1)^{\tilde{u}_{mj}} {\rm Pr}(\tilde{u}_{mj}=0)^{(1-\tilde{u}_{mj})}\\
&=\prod_{i=1}^{K} \prod_{j=1}^{N} {\rm Pr}(\tilde{u}_{\pi(i)j}=1)^{\tilde{u}_{\pi(i)j}} {\rm Pr}(\tilde{u}_{\pi(i)j}=0)^{(1-\tilde{u}_{\pi(i)j})},
\end{split}
\end{equation}
where $({\rm Pr}(\tilde{u}_{ij}=1),{\rm Pr}(\tilde{u}_{ij}=0))$ is the PMF of $\tilde{u}_{ij}$.
The corresponding log-likelihood function $l({\tilde {\boldsymbol \eta}};{\theta},{\boldsymbol \Pi})$ is
\begin{small}
\begin{align}\label{unlabel_log}
l({\tilde {\boldsymbol \eta}};{\theta},{\boldsymbol \Pi})=N\sum_{i=1}^{K} \left(\tilde\eta_{\pi(i)} \log p_i+(1-\tilde\eta_{\pi(i)})\log(1-p_i)\right),
\end{align}
\end{small}
where $\tilde \eta_{\pi(i)}=\sum_{j=1}^{N}{\tilde u}_{\pi(i)j}/N=\sum_{j=1}^{N}{\tilde u}_{mj}/N$. The ML estimation problem can be formulated as
\begin{equation}\label{unlabel_MLE}
\underset{\theta\in[\Delta,\Delta], {{\boldsymbol \Pi}\in {\mathcal P}_K}} {\operatorname{max}}~ l({\tilde {\boldsymbol \eta}};{\theta},{\boldsymbol \Pi}),
\end{equation}
where ${\mathcal P}_K$ denotes the set of all possible $K\times K$ permutation matrices.

\subsection{Estimation with permuted data and known $\theta$}\label{knownsubsec}
In this subsection, the permutation matrix recovery problem is studied in the case of known $\theta$. It is shown that the permutation matrix under ML estimation criterion can be recovered efficiently under the ML criterion.
\begin{proposition}\label{prop_1}
Given the ML estimation problem in (\ref{unlabel_MLE}) with known $\theta$, the ML estimate of the permutation matrix ${\boldsymbol \Pi}$ will reorder the rows of $\widetilde{\mathbf U}$, and equivalently the elements of $\tilde{\boldsymbol \eta}$, to have the same relative order as the elements of $(1-q_0-q_1)({\mathbf h}\theta-{\boldsymbol \tau})$.
\end{proposition}
\begin{IEEEproof}
Note that the objective function $l({\tilde {\boldsymbol \eta}};{\theta},{\boldsymbol \Pi})$ (\ref{unlabel_log}) can be decomposed as
\begin{align}\label{simpllh}
l({\tilde {\boldsymbol \eta}};{\theta},{\boldsymbol \Pi}) = K\sum_{i=1}^{N} {\tilde\eta_{\pi(i)}} s_i+K\sum_{i=1}^{N}\log(1-p_i),
\end{align}
where $ s_i=\log( p_i/(1-p_i))$.
From (\ref{simpllh}), the ML estimate of the permutation matrix ${\boldsymbol \Pi}$ will reorder the rows of $\mathbf U$, and equivalently the elements of $\tilde{\boldsymbol\eta}$ to have the same relative order as the elements of $\mathbf s$ \cite{Marano1, Jiang4}. Because $s_i$ is monotonically increasing with respect to $(1-q_0-q_1)(h_i\theta-\tau_i)$, the elements of $\tilde{\boldsymbol \eta}$ should be reordered by the permutation matrix to have the same relative order as the elements of $(1-q_0-q_1)({\mathbf h}\theta-{\boldsymbol \tau})$ to maximize the likelihood.
\end{IEEEproof}

If $\tau_i=c_0 h_i$, then changing $\theta-c_0$ in $-(\theta-c_0)$ would reverse the ordering. This might help to explain why two solutions appear in the subsequent Proposition \ref{statid} when $\theta$ is unknown.
\subsection{Estimation with permuted data and unknown $\theta$}
In general, $\theta$ may be unknown. Consequently, we should jointly estimate $\theta$ and permutation matrix $\boldsymbol \Pi$. However, finding the best permutation matrix is very challenging in most problems due to non-convexity. One could try all the possible permutation matrices, at complexity cost $O(N!)$. Given a permutation matrix, one obtains the ML estimate of $\theta$ via numerical algorithms and achieves global optimum under $q_0=q_1=0$ or $1$. Under $0<q_0+q_1<2$, we do not know
whether the negative log-likelihood function is convex or not, and local optimum is guaranteed. Given $\theta$, the computation complexity of finding the optimal permutation matrix is just reordering, which costs $O(N\log N)$, as we show in subsection \ref{knownsubsec}.
\subsubsection{Alternating maximization algorithm for general case}\label{subsub1}
The problem structure induces us to optimize the two unknowns alternately as shown in Algorithm \ref{alternating_m}.
\begin{algorithm}
\caption{ Alternating Maximization }
\begin{algorithmic}[1]\label{alternating_m}
\STATE Initialize $t=1$ and $\hat{\theta}_{t-1}$;
\STATE Fix ${\theta}=\hat{\theta}_{t-1}$, reorder $\tilde{\boldsymbol\eta}$ according to $(1-q_0-q_1)({\mathbf h}\theta-{\boldsymbol \tau})$ and obtain the corresponding permutation matrix $\hat{\boldsymbol\Pi}_{t-1}$;
\STATE Solve $\underset{\theta}{\operatorname{max}}~l({\tilde {\boldsymbol \eta}};{\theta},\hat{\boldsymbol\Pi}_{t-1})$ and obtain $\hat{\theta}_{t}$;
\STATE Set $t=t+1$ and return to step 2 until a sufficient number of iterations has been performed or $|\hat{\theta}_{t}-\hat{\theta}_{t-1}|\leq\epsilon$, where $\epsilon$ is a tolerance parameter.
\end{algorithmic}
\end{algorithm}
The alternating maximization in Algorithm \ref{alternating_m} can be viewed as the alternating projection with respect to $\theta$ and $\boldsymbol\Pi$. The objective function is $l({\tilde {\boldsymbol \eta}};{\theta},{\boldsymbol \Pi})$. In step $2$, given $\hat\theta_{t-1}$, we update the permutation matrix as $\hat{\boldsymbol\Pi}_{t-1}$, and the objective value is $l({\tilde {\boldsymbol \eta}};{\hat{\theta}_{t-1}},\hat{\boldsymbol\Pi}_{t-1})$. Given $\hat{\boldsymbol\Pi}_{t-1}$, we obtain ML estimation of $\theta$ as $\hat{\theta}_{t}$, and the objective value is $l({\tilde {\boldsymbol \eta}};{\hat{\theta}_{t}},\hat{\boldsymbol\Pi}_{t-1})$ satisfying $l({\tilde {\boldsymbol \eta}};{\hat{\theta}_{t}},\hat{\boldsymbol\Pi}_{t-1}) \geq l({\tilde {\boldsymbol \eta}};{\hat{\theta}_{t-1}},\hat{\boldsymbol\Pi}_{t-1})$. Given $\hat{\theta}_{t}$, we update the permutation matrix as $\hat{\boldsymbol\Pi}_{t}$, and the objective value is $l({\tilde {\boldsymbol \eta}};{\hat{\theta}_{t}},\hat{\boldsymbol\Pi}_{t})$ satisfying $l({\tilde {\boldsymbol \eta}};{\hat{\theta}_{t}},\hat{\boldsymbol\Pi}_{t})\geq l({\tilde {\boldsymbol \eta}};{\hat{\theta}_{t}},\hat{\boldsymbol\Pi}_{t-1})$. Consequently, we have
\begin{align}
l({\tilde {\boldsymbol \eta}};{\hat{\theta}_{t}},\hat{\boldsymbol\Pi}_{t}) \geq l({\tilde {\boldsymbol \eta}};{\hat{\theta}_{t-1}},\hat{\boldsymbol\Pi}_{t-1}).
\end{align}
Provided that the maximum with respect to each $\theta$ and $\boldsymbol \Pi$ is unique, any accumulation point of the sequence generated by Algorithm \ref{alternating_m} is a stationary point \cite{Bertsekas}.
\subsubsection{Special cases for efficient recovery of $\boldsymbol \Pi$ under unknown $\theta$}
\begin{proposition}\label{simplesort}
Given the ML estimation problem in (\ref{unlabel_MLE}) with unknown $\theta$, if there exist constants $c, d, e\in{\mathbb R}$ such that $c{\boldsymbol \tau}+d{\mathbf h}=e{\mathbf 1}$, the elements of $\tilde{\boldsymbol \eta}$ should be reordered according to the order of the elements of $(q_0+q_1-1){\boldsymbol \tau} $ if $c=0$, otherwise reordered according to ${\mathbf h}$ or $-{\mathbf h}$.
\end{proposition}
\begin{IEEEproof}
We separately address the cases $c=0$ and $c\neq 0$. In the case of $c=0$, $\mathbf h$ must be a constant vector. Reordering according to $(1-q_0-q_1)({\mathbf h}\theta-{\boldsymbol \tau})$ is equivalent to reordering according to $(q_0+q_1-1){\boldsymbol \tau}$. Since $(q_0,q_1)$ are known in this problem, $\tilde{\boldsymbol \eta}$ should be reordered according to ${\boldsymbol \tau}$ if $q_0+q_1>1$ or $-{\boldsymbol \tau}$ if $q_0+q_1<1$. In the case of $c\neq 0$, we have ${\boldsymbol \tau}=(e/c){\mathbf 1}-(d/c){\mathbf h}$. Consequently, ${\mathbf h}\theta-{\boldsymbol \tau}=(\theta+d/c){\mathbf h}-(e/c){\mathbf 1}$, and $\tilde{\boldsymbol \eta}$ is reordered according to ${\mathbf h}$ or $-{\mathbf h}$.
\end{IEEEproof}

The above proposition deals with four cases, i.e., $\mathbf h$ is a constant vector $(c=0)$, $\boldsymbol \tau$ is a constant vector $(d=0)$, $\mathbf h$ is a multiple of $\boldsymbol \tau$ $(e=0)$ and each pair of components of $\mathbf h$ and $\boldsymbol \tau$ lies in the same line $c\tau_i+d h_i=e$ $(cde\not=0)$. In \cite{Jiang4} it is shown that reordering yields the optimal MLE given ${\mathbf h}={\mathbf 1}$. Proposition \ref{simplesort} extends the special case in \cite{Jiang4} to more general cases. Consequently, we propose Algorithm \ref{reorder}, an efficient algorithm for parameter estimation.
\begin{algorithm}
\caption{ Reordering algorithm }
\begin{algorithmic}[1]\label{reorder}
\STATE If $c=0$, reorder the elements of $\tilde{\boldsymbol \eta}$ according to the elements of $(q_0+q_1-1)\boldsymbol \tau$. The corresponding permutation matrix is $\hat{\boldsymbol\Pi}_{s0}$. Solve the parameter estimation problem by numerical algorithm and obtain $\hat{\theta}_{\rm ML}=\underset{\theta}{\operatorname{argmax}}~ l({\tilde {\boldsymbol \eta}};{\theta},\hat{\boldsymbol\Pi}_{s0})$;
\STATE If $c\not=0$, reorder the elements of $\tilde{\boldsymbol \eta}$ according to the elements of ${\mathbf h}$ and $-{\mathbf h}$. The corresponding permutation matrices are $\hat{\boldsymbol\Pi}_{s1}$ and $\hat{\boldsymbol\Pi}_{s2}$;
\STATE Solve the single variable optimization problems and obtain $\hat{\theta}_{s1}=\underset{\theta}{\operatorname{argmax}}~ l({\tilde {\boldsymbol \eta}};{\theta},\hat{\boldsymbol\Pi}_{s1})$ and $\hat{\theta}_{s2}=\underset{\theta}{\operatorname{argmax}}~ l({\tilde {\boldsymbol \eta}};{\theta},\hat{\boldsymbol\Pi}_{s2})$. Choose $\hat{\theta}_{\rm ML}=\hat{\theta}_{s1}$ given that $l({\tilde {\boldsymbol \eta}};\hat{\theta}_{s1},\hat{\boldsymbol\Pi}_{s1}) \geq l({\tilde {\boldsymbol \eta}};\hat{\theta}_{s2},\hat{\boldsymbol\Pi}_{s2})$, otherwise $\hat{\theta}_{\rm ML}=\hat{\theta}_{s2}$.
\end{algorithmic}
\end{algorithm}
\subsection{Statistical identifiability}
Note that Algorithm \ref{reorder} may generate two solutions $(\hat{\theta}_{s1},\hat{\boldsymbol\Pi}_{s1})$ and $(\hat{\theta}_{s2},\hat{\boldsymbol\Pi}_{s2})$.
Given system parameters $\mathbf h$ and $\boldsymbol \tau$, it is important to determine whether the two solutions $(\hat{\theta}_{s1},\hat{\boldsymbol\Pi}_{s1})$ and $(\hat{\theta}_{s2},\hat{\boldsymbol\Pi}_{s2})$ will yield the same log-likelihood $l({\tilde {\boldsymbol \eta}};\hat{\theta}_{s1},\hat{\boldsymbol\Pi}_{s1}) = l({\tilde {\boldsymbol \eta}};\hat{\theta}_{s2},\hat{\boldsymbol\Pi}_{s2})$. If $l({\tilde {\boldsymbol \eta}};\hat{\theta}_{s1},\hat{\boldsymbol\Pi}_{s1}) = l({\tilde {\boldsymbol \eta}};\hat{\theta}_{s2},\hat{\boldsymbol\Pi}_{s2})$, two pairs of parameter values lead to the same maximum likelihood. In this situation, $(\theta,{\boldsymbol\Pi})$ clearly cannot be estimated consistently since $\tilde{\boldsymbol \eta}$ provide no information as to whether it is $(\hat{\theta}_{s1},\hat{\boldsymbol\Pi}_{s1})$ or $(\hat{\theta}_{s2},\hat{\boldsymbol\Pi}_{s2})$. This phenomenon motivates us delving into the identifiability of the model. Statistical identifiability is a property of a statistical model which describes one-to-one correspondence between parameters and probability distributions \cite{Lehmann1}. In this subsection, we provide the following proposition which justifies that there exist cases in which the model is unidentifiable, i.e., there exist two different parameter values ($({\theta}_{s1},{\boldsymbol\Pi}_{s1})$ and $({\theta}_{s2},{\boldsymbol\Pi}_{s2})$) leading to the same distribution of the observations ${\tilde {\boldsymbol \eta}}$ \cite{Lehmann1}.
\begin{proposition}\label{statid}
Let ${\mathbf h}_a$ and ${\mathbf h}_d$ denote the ascending and descending ordered versions of $\mathbf h$, and ${\boldsymbol \Pi}_a{\mathbf h}={\mathbf h}_a$ and ${\boldsymbol \Pi}_d{\mathbf h}={\mathbf h}_d$, where ${\boldsymbol \Pi}_a$ and ${\boldsymbol \Pi}_d$ are permutation matrices. Given ${\boldsymbol \tau}=c_0{\mathbf h}$ and ${\mathbf h}_a=-{\mathbf h}_d$, the model is unidentifiable, i.e., $l({\tilde {\boldsymbol \eta}};{\theta},{\boldsymbol \Pi})|_{\theta={\theta}_{s1},{\boldsymbol \Pi}={\boldsymbol\Pi}_{s1}} =l({\tilde {\boldsymbol \eta}};{\theta},{\boldsymbol \Pi})|_{\theta={\theta}_{s2},{\boldsymbol \Pi}={\boldsymbol\Pi}_{s2}}$, where ${\theta}_{s2}=2c_0-{\theta}_{s1}$ and ${\boldsymbol\Pi}_{s2}={\boldsymbol\Pi}_{s1}{\boldsymbol \Pi}_a^{\rm T}{\boldsymbol \Pi}_d$.
\end{proposition}
\begin{IEEEproof}
Let ${\boldsymbol\Pi}_{s1}$ be a permutation matrix such that ${\boldsymbol\Pi}_{s1}^{\rm T}{\tilde {\boldsymbol \eta}}$ has the same relative order as $\mathbf h$. Now we prove that ${\boldsymbol\Pi}_{s2}^{\rm T}{\tilde {\boldsymbol \eta}}$ has the same relative order as $-\mathbf h$. Utilizing ${\mathbf h}_a=-{\mathbf h}_d=-{\boldsymbol \Pi}_d{\mathbf h}$ and ${\boldsymbol \Pi}_d{\boldsymbol \Pi}_d^{\rm T}={\mathbf I}$, we obtain ${\boldsymbol \Pi}_d^{\rm T}{\mathbf h}_a=-{\mathbf h}$. Note that ${\boldsymbol\Pi}_{s2}^{\rm T}{\tilde {\boldsymbol \eta}}={\boldsymbol \Pi}_d^{\rm T}{\boldsymbol \Pi}_a{\boldsymbol\Pi}_{s1}^{\rm T}{\tilde {\boldsymbol \eta}}$. Because ${\boldsymbol\Pi}_{s1}^{\rm T}{\tilde {\boldsymbol \eta}}$ has the same relative order as $\mathbf h$, ${\boldsymbol \Pi}_a{\boldsymbol\Pi}_{s1}^{\rm T}{\tilde {\boldsymbol \eta}}$ has the same relative order as ${\boldsymbol \Pi}_a\mathbf h={\mathbf h}_a$, and ${\boldsymbol \Pi}_d^{\rm T}{\boldsymbol \Pi}_a{\boldsymbol\Pi}_{s1}^{\rm T}{\tilde {\boldsymbol \eta}}$ has the same relative order as ${\boldsymbol \Pi}_d^{\rm T}{\mathbf h}_a=-{\mathbf h}$.

Next we prove that $l({\tilde {\boldsymbol \eta}};{\theta},{\boldsymbol \Pi})|_{\theta={\theta}_{s1},{\boldsymbol \Pi}={\boldsymbol\Pi}_{s1}} =l({\tilde {\boldsymbol \eta}};{\theta},{\boldsymbol \Pi})|_{\theta={\theta}_{s2},{\boldsymbol \Pi}={\boldsymbol\Pi}_{s2}}$ holds. Because ${\theta}_{s2}=2c_0-{\theta}_{s1}$, we have
\begin{equation}\label{sim1hasym}
h_i{\theta}_{s1}-\tau_i=h_i({\theta}_{s1}-c_0),\quad h_i{\theta}_{s2}-\tau_i=-h_i({\theta}_{s1}-c_0).
\end{equation}
By examining $l({\tilde {\boldsymbol \eta}};{\theta},{\boldsymbol \Pi})$ (\ref{simpllh}) and utilizing ${\mathbf h}_a=-{\mathbf h}_d$, the second term of $l({\tilde {\boldsymbol \eta}};{\theta},{\boldsymbol \Pi})|_{\theta={\theta}_{s1},{\boldsymbol \Pi}={\boldsymbol\Pi}_{s1}}$ is equal to that of $l({\tilde {\boldsymbol \eta}};{\theta},{\boldsymbol \Pi})|_{\theta={\theta}_{s2},{\boldsymbol \Pi}={\boldsymbol\Pi}_{s2}}$. For the first term, note that given ${\theta}_{s1}$ and ${\theta}_{s2}$, the corresponding ${\mathbf s}_1$ and ${\mathbf s}_2$ in (\ref{simpllh}) can be viewed as evaluating at $\mathbf h$ and $-\mathbf h$ according to (\ref{sim1hasym}), respectively. Because ${\mathbf h}_a=-{\mathbf h}_d$, we can conclude that ${\mathbf s}_1$ is a permutated version of ${\mathbf s}_2$. The first term of (\ref{simpllh}) can be expressed as either $({\boldsymbol\Pi}_{s1}^{\rm T}{\tilde {\boldsymbol \eta}})^{\rm T}{\mathbf s}_1$ or $({\boldsymbol\Pi}_{s2}^{\rm T}{\tilde {\boldsymbol \eta}})^{\rm T}{\mathbf s}_2$. Because $({\boldsymbol\Pi}_{s1}^{\rm T}{\tilde {\boldsymbol \eta}})^{\rm T}$ and ${\mathbf s}_1$ have the same relative order as $\mathbf h$, and $({\boldsymbol\Pi}_{s2}^{\rm T}{\tilde {\boldsymbol \eta}})^{\rm T}$ and ${\mathbf s}_2$ have the same relative order as $-\mathbf h$, one has $({\boldsymbol\Pi}_{s1}^{\rm T}{\tilde {\boldsymbol \eta}})^{\rm T}{\mathbf s}_1=({\boldsymbol\Pi}_{s2}^{\rm T}{\tilde {\boldsymbol \eta}})^{\rm T}{\mathbf s}_2$. Thus $l({\tilde {\boldsymbol \eta}};{\theta},{\boldsymbol \Pi})|_{\theta={\theta}_{s1},{\boldsymbol \Pi}={\boldsymbol\Pi}_{s1}}=l({\tilde {\boldsymbol \eta}};{\theta},{\boldsymbol \Pi})|_{\theta={\theta}_{s2},{\boldsymbol \Pi}={\boldsymbol\Pi}_{s2}}$.
\end{IEEEproof}

Now an example is presented to substantiate the above
proposition. Let $c_0=0.5$, the true value $\theta_0=1$, ${\mathbf h}=[2,-1,-2,1]^{\rm T}$, ${\boldsymbol \eta}=[\eta_1,\eta_2,\eta_3,\eta_4]^{\rm T}$ and ${\boldsymbol \Pi}_0=[0~ 0~ 1 ~0;0 ~1~ 0~ 0;0 ~0 ~0~ 1;1~ 0 ~0 ~0]$. Then $\tilde{\boldsymbol \eta}=[\eta_3,\eta_2,\eta_4,\eta_1]^{\rm T}$, ${\mathbf h}_a=[-2,-1,1,2]^{\rm T}$, ${\mathbf h}_d=[2,1,-1,-2]^{\rm T}$ and ${\mathbf h}_a=-{\mathbf h}_d$. We can conclude that $l({\tilde {\boldsymbol \eta}};{\theta},{\boldsymbol \Pi})|_{\theta=1,{\boldsymbol \Pi}={\boldsymbol\Pi}_0}=l({\tilde {\boldsymbol \eta}};{\theta},{\boldsymbol \Pi})|_{\theta=0,{\boldsymbol \Pi}={\boldsymbol\Pi}'}$, where ${\boldsymbol \Pi}'=[0~ 0 ~ 0~ 1;1 ~0~ 0~ 0;0~ 0 ~1~ 0;0~ 1~ 0 ~0]$.

In addition,  given $|c_0|\geq \Delta$, only one of $\{\theta_{s1},\theta_{s2}\}$ lies in the interval $[-\Delta,\Delta]$, and the model is identifiable. In the following, a method to select good initial points for the alternating maximization algorithm is provided for the general case.
\subsection{Good initial points}
For alternating maximization algorithms dealing with nonconvex optimization problems, an initial point is important for the algorithm to converge to the global optimum. In the following text, we provide good initial points for the alternating maximization algorithm. The key idea is to obtain a coarse estimate of $\theta$ via matching the expected and actual number of ones in observations, and utilizing the orthogonal property of permutation matrix.

Suppose that the number of measurements $K$ is large. Consequently, as the number of measurements tends to infinity, the law of large numbers (LLN) implies

\begin{align}\label{LLN}
\eta_i\overset{\rm p}\longrightarrow q_0+(1-q_0-q_1)F_w\left(({h_i{\theta}-\tau_i})/{\sigma_w}\right),
\end{align}
where $\overset{\rm p}\longrightarrow$ denotes convergence in probability.
Given $\theta \in [-\Delta,\Delta]$, $-|h_i|\Delta-\tau_i\leq h_i\theta-\tau_i\leq  |h_i|\Delta-\tau_i$.
In the following text, we only deal with $q_0+q_1<1$ case. The case that $q_0+q_1>1$ is very similar and is omitted here.
Define $l=\underset{i\in [1,\cdots,N]}{\operatorname{min}}~\left(q_0+(1-q_0-q_1)F_w(({-|h_i|\Delta-\tau_i})/{\sigma_w})\right)$ and $u=\underset{i\in [1,\cdots,N]}{\operatorname{max}}~\left(q_0+(1-q_0-q_1)F_w(({|h_i|\Delta-\tau_i})/{\sigma_w})\right)$.
Then ${\eta}_i$ should satisfy $l\leq {\eta}_i \leq u$. Let ${\mathcal I}_{l,u}(\tilde\eta_i)$ denotes the projection of $\tilde\eta_i$ onto the interval $[l,u]$. Note that this projection operation is needed because (\ref{LLN}) is valid in the limit as $K$ goes to infinity.
From (\ref{LLN}) one obtains
\begin{small}
\begin{align}
{\mathbf m}\triangleq  \sigma_wF_w^{-1}\left(({{{\mathcal I}_{l,u}(\tilde{\boldsymbol\eta})-q_0{\mathbf 1}_N}})/({{1-q_0-q_1}})\right) {\overset{\rm p}\longrightarrow}{\boldsymbol \Pi}({\mathbf h}{\theta}-{\boldsymbol \tau}) . \notag
\end{align}
\end{small}
Utilizing ${\boldsymbol \Pi}{\boldsymbol \Pi}^{\rm T}=\mathbf I$ yields
\begin{align}\label{mt_m}
{\mathbf m}^{\rm T}{\mathbf m}{\overset{\rm p}\longrightarrow} {\mathbf h}^{\rm T}{\mathbf h}{\theta}^2-2{\boldsymbol \tau}^{\rm T}{\mathbf h}{\theta}+{\boldsymbol \tau}^{\rm T}{\boldsymbol \tau},
\end{align}
which is a quadratic equation in $\theta$. Accordingly, using the asymptotic properties of ${\mathbf m}^{\rm T}{\mathbf m}$, one obtains (\ref{goodinits}) via inverting (\ref{mt_m})
\begin{align}\label{goodinits}
\theta_{1,2}=\frac{{\boldsymbol\tau}^{\rm T}{\mathbf h}}{{\mathbf h}^{\rm T}{\mathbf h}}\pm
\sqrt{\frac{{\mathbf m}^{\rm T}{\mathbf m}-{\boldsymbol\tau}^{\rm T}{\boldsymbol\tau}}{{\mathbf h}^{\rm T}{\mathbf h}}
+(\frac{{\boldsymbol\tau}^{\rm T}{\mathbf h}}{{\mathbf h}^{\rm T}{\mathbf h}})^2}.
\end{align}
The above two solutions can be used for the alternating maximization algorithm as initial points. Finally, the optimum with larger likelihood is chosen as ML estimator. In Section \ref{simulation}, to provide a fair comparison of the alternating maximization algorithm with good initial points, $-\Delta$ and $\Delta$ are used as two initial points, and we choose as ML estimator the solution whose likelihood is larger.

The result of (\ref{goodinits}) is consistent with that of Proposition \ref{statid}. Given that the conditions in Proposition \ref{statid} are satisfied, and substituting ${\boldsymbol \tau}=c_0{\mathbf h}$ into (\ref{goodinits}), the solutions are $\theta_1=\theta$ and $\theta_2=2c_0-\theta$.
\section{Detection from unlabeled data}\label{det}
In this section, we study the detection problem from unlabeled data. The GLRT detector is studied separately in the cases of known and unknown $\theta$. In addition, we investigate the permutation matrix recovery probability.
\subsection{Detection with permuted data and known $\theta$}\label{GLRT_T2}
In the case of known $\theta$, the GLRT can be formulated as
\begin{equation}
\begin{aligned}\label{glrtknowntheta}
T_2(\tilde{\boldsymbol\eta})={\underset{{\boldsymbol \Pi}\in {\mathcal P}_N} {\operatorname{max}}~ l({\tilde {\boldsymbol \eta}};{\theta},{\boldsymbol \Pi})}-
{ \underset{{\boldsymbol \Pi}\in {\mathcal P}_N} {\operatorname{max}}~l({\tilde {\boldsymbol \eta}};0,{\boldsymbol \Pi})}>\gamma.
\end{aligned}
\end{equation}
As shown in Proposition \ref{prop_1}, the ML estimate of the permutation matrix ${\boldsymbol \Pi}$ corresponding to the first term in (\ref{glrtknowntheta}) will reorder the elements of $\tilde{\boldsymbol\eta}$ to have the same relative order as the elements of $(1-q_0-q_1)({\mathbf h}\theta-{\boldsymbol \tau})$. Similarly, for the ML estimation problem corresponding to the second term in (\ref{glrtknowntheta}), we reorder the elements of $\tilde{\boldsymbol\eta}$ to have the same order as that of $-(1-q_0-q_1){\boldsymbol \tau}$.
\subsection{Detection with permuted data and unknown $\theta$}\label{GLRT_T3}
For the unknown $\theta$ and unknown $\boldsymbol \Pi$ case, a GLRT is used to decide ${\mathcal H}_1$ if
\begin{align}\label{GLRTjoint}
T_3(\tilde{\boldsymbol\eta})={\underset{\theta, {{\boldsymbol \Pi}\in {\mathcal P}_N}} {\operatorname{max}}~ l({\tilde {\boldsymbol \eta}};{\theta},{\boldsymbol \Pi})}-
{\underset{{\boldsymbol \Pi}\in {\mathcal P}_N} {\operatorname{max}}~ l({\tilde {\boldsymbol \eta}};0,{\boldsymbol \Pi})}>\gamma.
\end{align}
Algorithm \ref{alternating_m} for joint estimation of $\theta$ and $\boldsymbol \Pi$ has been described in section \ref{subsub1}. The performance of the GLRT (\ref{GLRTjoint}) will be evaluated in the Algorithm 1 for joint estimation of $\theta$ and $\boldsymbol \Pi$ is
necessary for the first term and has been described in section \ref{subsub1}; the second term is as in section \ref{GLRT_T2}.
\subsection{Approximations on permutation matrix recovery probability}\label{bounds_l_u}
In this subsection, we investigate the permutation matrix recovery probability problem. Since errors in permutation
matrix recovery are more likely to happen in the relatively indistinguishable cases, the performances in terms of signal detection or estimation tasks may not be closely related to the recovery of permutation matrix. However, it is meaningful to extract the accurate timestamp information or sensors' identity information which corresponds to recovery of permutation matrix, as presented in the following.

It is difficult to obtain the permutation matrix recovery probability in the case of unknown $\theta$. Instead, we assume that $\theta$ is known, and analyze the permutation matrix recovery probability in terms of the recovery algorithm provided in Proposition \ref{prop_1}. Without loss of generality, we also assume that $q_0+q_1<1$ in the following analysis. The case that $q_0+q_1>1$ is similar and is omitted here.

First, let $p_i$ be ordered such that $p_{(1)}>p_{(2)}> \cdots> p_{(K)}$. From (\ref{y_probability}) we have $(h_i\theta-\tau_i)_{(1)}>(h_i\theta-\tau_i)_{(2)}>\cdots> (h_i\theta-\tau_i)_{(K)}$. Provided $q_0+q_1<1$, the elements of $\tilde{\boldsymbol\eta}$ should be reordered according to the order of the elements of $\mathbf h{\theta}-\boldsymbol \tau$ in Proposition \ref{prop_1}. 
Therefore the permutation matrix will be correctly recovered if and only if $\eta_{(1)} > \eta_{(2)} >\cdots> \eta_{(K)}$. Note that the subscripts of $(h_i\theta-\tau_i)_{(\cdot)}$ and $\eta_{(\cdot)}$ also correspond to the order of $p_i$, instead of the order of $h_i\theta-\tau_i$ or $\eta_i$.

Define $E_i$ as the event such that $\eta_{(i)}>\eta_{(i+1)}$ and $\bar{E_i}$ as the corresponding complement event of $E_i$, namely, $\eta_{(i)}\leq\eta_{(i+1)}$. The probability that permutation matrix is recovered correctly can be written as
\begin{equation}
\begin{split}\label{unionbound}
&{\rm Pr}(\hat{\boldsymbol \Pi}_{\rm ML}={\boldsymbol \Pi}_0)={\rm Pr}(\eta_{(1)} >\cdots> \eta_{(K)})={\rm Pr}\left(\bigcap\limits_{i=1}^{K-1} E_i\right)\notag\\
=&1-{\rm Pr}\left(\bigcup\limits_{i=1}^{K-1} {\bar{E}_i}\right)\geq 1-\sum\limits_{i=1}^{K-1}{\rm Pr}(\bar{E}_i),
\end{split}
\end{equation}
where union bound ${\rm Pr}\left(\bigcup\limits_{i=1}^{K-1} {\bar{E}_i}\right)\leq \sum\limits_{i=1}^{K-1}{\rm Pr}(\bar{E}_i)$ is utilized in (\ref{unionbound}). From (\ref{y_probability}), we have $u_{ij}\sim\mathcal B (1, p_i)$ and $N\eta_i=\sum_{j=1}^{N} u_{ij} \sim \mathcal B(N, p_i)$. When $N$ is large, the De Moivre-Laplace~theorem \cite{Papoulis1} implies that the distribution of $\eta_i$ can be approximated by ${\mathcal N}(p_i,{p_i(1-p_i)/N})$. As a consequence, $\eta_{(i)}-\eta_{(i+1)}$ is approximately distributed as ${\mathcal N}(p_{(i)}-p_{(i+1)},p_{(i)}(1-p_{(i)})/N+p_{(i+1)}(1-p_{(i+1)})/N)$, and
\begin{align}
&{\rm Pr}(\hat{\boldsymbol \Pi}_{\rm ML}={\boldsymbol \Pi}_0) \geq 1-\sum\limits_{i=1}^{K-1}{\rm Pr}(\bar{E}_i)= 1-\sum\limits_{i=1}^{K-1}{\rm Pr}({\eta_{(i)}-\eta_{(i+1)}}\leq 0)\notag\\
\approx & 1-\sum\limits_{i=1}^{K-1} {\Phi\left(\frac{-(p_{(i)}-p_{(i+1)})\sqrt{N}}{\sqrt{p_{(i)}(1-p_{(i)})+p_{(i+1)}(1-p_{(i+1)})}}\right)}\notag\\
\geq & 1-(K-1)\Phi\left(-t\sqrt{N}\right)\notag\\
\approx &1-(K-1)\frac{1}{\sqrt{2\pi}t\sqrt{N}} e^{-t^2 N/2}\notag\\
= &1-\frac{1}{\sqrt{2\pi}} e^{\ln (K-1)-\ln t-\frac{1}{2}\ln N-\frac{t^2}{2} N}\triangleq {\rm Pr}(K,N),\label{theory_approx}
\end{align}
where $t=\underset{i=1,\cdots,K-1}{\operatorname{min}}~\frac{v_i}{\sqrt{p_{(i)}(1-p_{(i)})+p_{(i+1)}(1-p_{(i+1)})}}$, $v_i=p_{(i)}-p_{(i+1)}$ and the approximation $\Phi(-x)\approx \frac{1}{\sqrt{2\pi}x}{\rm e}^{-\frac{x^2}{2}} (x\gg 0)$ is utilized.

Utilizing $p_{(i)}(1-p_{(i)})+p_{(i+1)}(1-p_{(i+1)})\leq 1/2$, we define $\tilde t$ satisfying
\begin{align}\label{def_tl}
\tilde{t}=\underset{i=1,\cdots,K-1}{\operatorname{min}}~{v_i}\leq \frac{\sqrt{2}}{2} t.
\end{align}

We conjecture that $\tilde t$ is on the order of $K^{-\alpha}$, i.e., $\tilde t=O(K^{-\alpha})$, which means that there exists constant $c_t$ such that
\begin{align}\label{conjecture_t}
\tilde{t} \approx c_t K^{-\alpha}.
\end{align}
In the following text, we show that we can construct $\mathbf h$ such that $\tilde{t}=O( K^{-1})$ and $\tilde{t}=O( K^{-2})$. According to (\ref{def_tl}) and (\ref{conjecture_t}), the approximation ${\rm Pr}(K,N)$ (\ref{theory_approx}) can be further simplified and relaxed as
\begin{align}
\widetilde{\rm Pr}(K,N)&=1-\frac{1}{2\sqrt{\pi}} e^{\ln (K-1)-\ln {\tilde t}-\frac{1}{2}\ln N-{{\tilde t}^2} N}\label{theory_NK1}\\
&\approx 1-\frac{1}{2\sqrt{\pi}c_t } e^{(1+\alpha)\ln K-\frac{1}{2}\ln N-\frac{c_t^2}{K^{2\alpha}} N}.\label{theory_NK2}
\end{align}
From (\ref{theory_NK2}), the exponent $(1+\alpha)\ln K-\frac{1}{2}\ln N-\frac{c_t^2}{K^{2\alpha}} N$ of (\ref{theory_NK2}) must be far less than $0$ for the recovery of permutation matrix. Given $N$ is large, the term $-\frac{1}{2}\ln N$ is small compared to $N$. Thus $(1+\alpha)\ln K-\frac{c_t^2}{K^{2\alpha}} N<0$ will ensure that the permutation matrix can be recovered in high probability. Simplifying $(1+\alpha)\ln K-\frac{c_t^2}{K^{2\alpha}} N<0$ yields
\begin{align}\label{simbound}
N>\frac{(1+\alpha)}{c_t^2}K^{2\alpha}\ln K.
\end{align}

The following cases are examples to illustrate $\tilde{t}=O(K^{-\alpha})$. For simplicity, we assume $1-q_0-q_1>0$, $\boldsymbol\tau = c\mathbf h (c<\theta)$ and
\begin{align}\label{adef}
a\triangleq (\theta-c)/{\sigma_w}>0.
\end{align}
\subsubsection{$\tilde{t}=O(K^{-1})$}
Let $\mathbf h$ be the shape of a ramp signal such that $h_i = u-\frac{(u-l)(i-1)}{K-1} (u>|l|)$, and $w_{ij}\sim{\mathcal N}(0,\sigma_w^2)$. Then the ordered sequence $p_{(i)}=p_i$, and $\tilde t$ can be approximated as

\begin{equation}
\begin{split}
\tilde{t}=&\underset{i=1,\cdots,K-1}{\operatorname{min}}~{p_i-p_{i+1}}\\
=&\frac{a(1-q_0-q_1)(u-l)}{K-1}\underset{i=1,\cdots,K-1}{\operatorname{min}}~f_w(a\xi_{i})\\
\approx &\frac{a(1-q_0-q_1)(u-l)f_w(au)}{K-1}\\
\approx&c_t K^{-1},
\end{split}
\end{equation}
where mean value theorem is utilized for $\xi_{i} \in (h_{i+1},h_i)$, $\xi_1\approx h_1=u$ is utilized when $K$ is large, and
\begin{align}\label{c_t}
c_t=a(1-q_0-q_1)(u-l)f_w(au).
\end{align}
Therefore $\tilde{t}$ can be reshaped in the form of (\ref{conjecture_t}).
\subsubsection{$\tilde{t}=O( K^{-2})$}
\begin{figure}[h!t]
 \setlength{\belowcaptionskip}{-0.5 cm}
\centering
\includegraphics[width=3.4in]{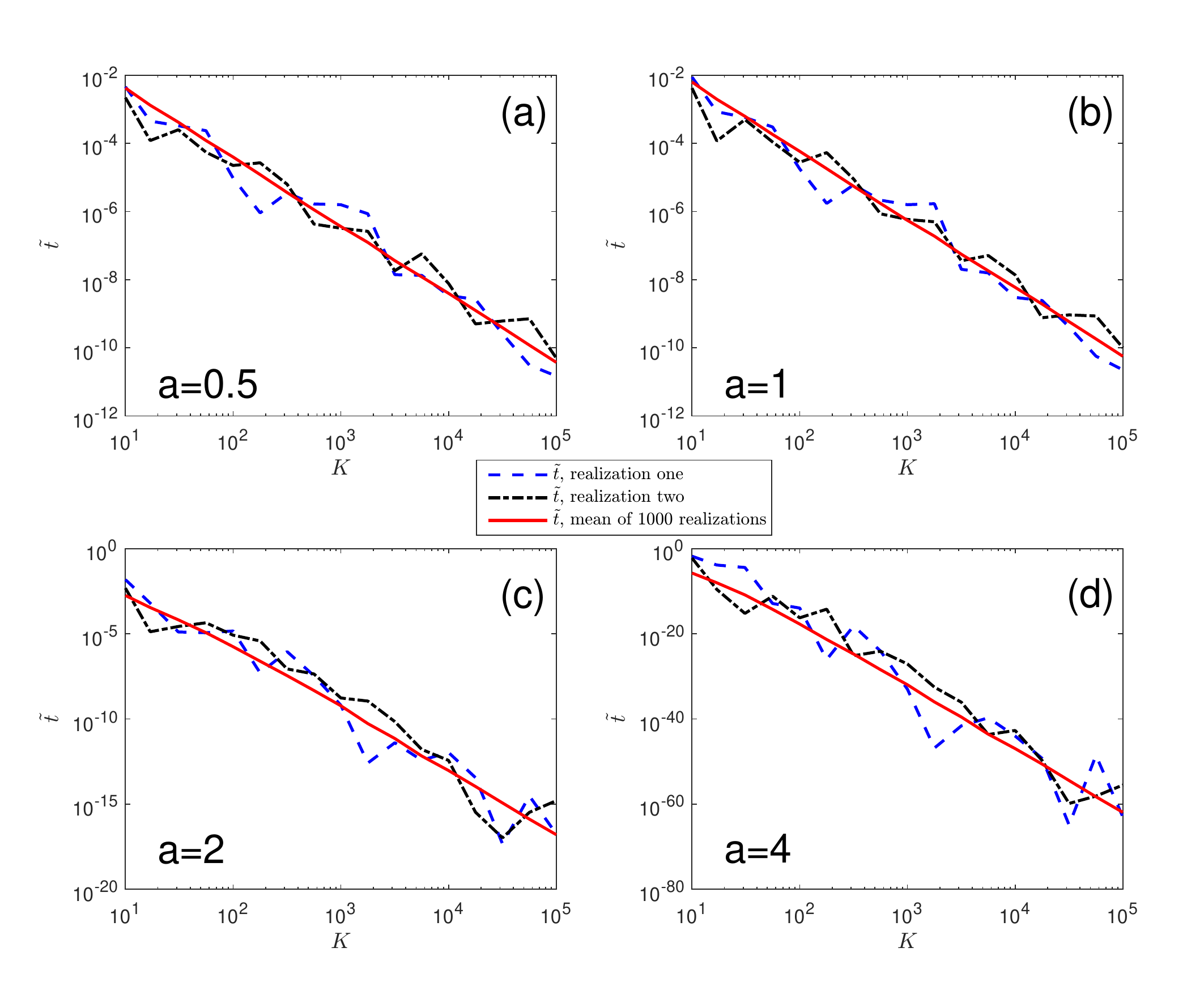}
\caption{The relationship of $\tilde t$ and $K$ under different $a$. Note that $q_0=q_1=0$, $h_i\sim{\mathcal N}(0,1)$ and $w_{ij}\sim{\mathcal N}(0,\sigma_w^2)$. }
\label{a_different}
\end{figure}

\begin{figure}[h!t]
\centering
\includegraphics[width=3.4in]{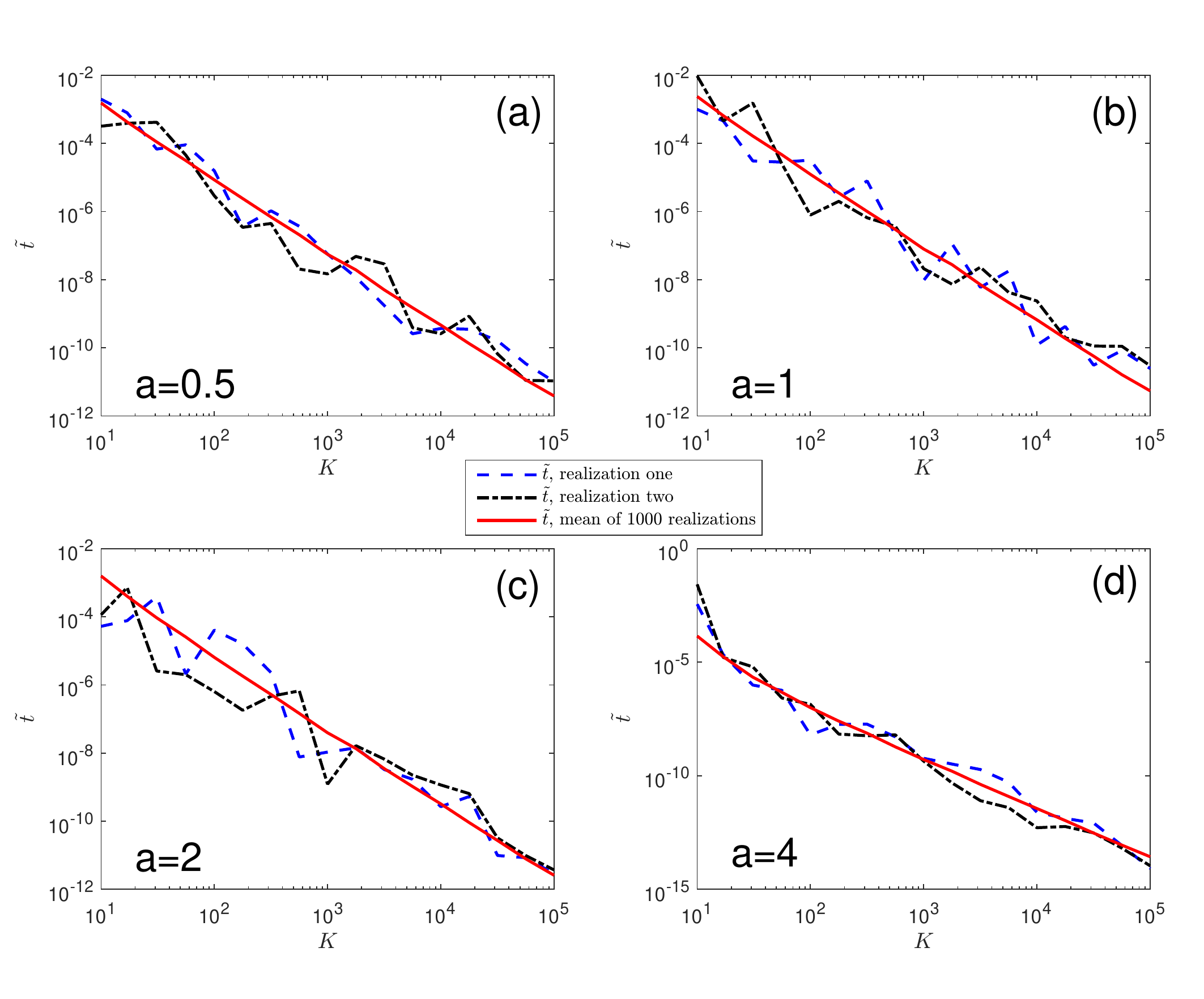}
\caption{The relationship of $\tilde t$ and $K$ under different $a$. Note that $q_0=q_1=0$, $h_i=\sin (2\pi x_i)$, $x_i\sim{\mathcal U}(0,1)$ and $w_{ij}\sim{\mathcal N}(0,\sigma_w^2)$.}
\label{t_K_sine}
\end{figure}

Let $h_i$ be independently drawn from the same distribution of $w_{ij}/\sigma_w$. The CDF of $p_i$ is
\begin{equation}
\begin{aligned}
&F_{p_i}(x)={\rm Pr}(p_i\leq x)\\
&={\rm Pr}(q_0+(1-q_0-q_1)F_w(ah_i)\leq x)\\
&={\rm Pr}\left(h_i\leq \frac{1}{a}F_w^{-1}\left(\frac{x-q_0}{1-q_0-q_1}\right)\right)\\
&=F_w\left(\frac{1}{a}F_w^{-1}\left(\frac{x-q_0}{1-q_0-q_1}\right)\right).
\end{aligned}
\end{equation}
In this case, we conjecture that $\tilde t = O(K^{g(a)})$, where $g(a)$ is a function of $a$, and the numerical results under different $a$ are shown in Fig. \ref{a_different}. In addition, the case in which $\mathbf h$ is the shape of a sinusoidal signal is also presented in Fig. \ref{t_K_sine}.

Now we prove that $\tilde{t}=O( K^{-2})$ under certain conditions. Given that $h_i$ and $w_{ij}/\sigma_w$ are i.i.d. random variables and $a=1$, the CDF $F_{p_i}(x)=(x-q_0)/(1-q_0-q_1)$, and the PDF of $p_i$ is
\begin{align}
f_{p_i}(x)=
\begin{cases}
&\frac{1}{1-q_0-q_1}, ~~q_0\leq x\leq 1-q_1,~~~~~~~~~\\
&0,\quad\quad {\rm otherwise}.
\end{cases}
\end{align}
Then the variates $p_{(1)}, p_{(2)}\cdots,p_{(K)}$ are distributed as $K$ descending ordered statistics from an uniform $(q_0,1-q_1)$ parent.
For $x\leq (1-q_0-q_1)/(K-1)$, the CDF of $\tilde{t}$ can be derived as \cite{David1} (page 135, equation (6.4.3))
\begin{equation}
\begin{aligned}
F_{\tilde{t}}(x)=&{\rm Pr}\left(\underset{i=1,\cdots,K-1}{\operatorname{min}}~{v_i}\leq x\right) \\
=&1-{\rm Pr}(v_1>x,v_2>x,\cdots,v_{K-1}>x)\\
=& 1- \left[1-\frac{(K-1)x}{1-q_0-q_1}\right]^K.
\end{aligned}
\end{equation}
For $x\geq (1-q_0-q_1)/(K-1)$, $F_{\tilde{t}}(x)=1$. Then the PDF of $\tilde{t}$ is
\begin{align}
f_{\tilde{t}}(x)=
\begin{cases}
&\!\!\!\frac{K(K-1)}{1-q_0-q_1}\left[1-\frac{(K-1)x}{1-q_0-q_1}\right]^{K-1}\!\!\!\!\!\!,~ 0\leq x\leq \frac{1-q_0-q_1}{K-1},\\
&\!\!\! 0, \qquad {{\rm otherwise}}.
\end{cases}
\end{align}
The expectation of $\tilde{t}$ is
\begin{equation}\label{t_expectation}
\begin{aligned}
{\rm E}_{\tilde{t}}[\tilde{t}]&=\int_0^1 x f_{\tilde{t}}(x)~{\rm d}x= \int_0^{\frac{1-q_0-q_1}{K-1}} \tilde{t} f_{\tilde{t}}(x)~{\rm d}x\\
&= \frac{K(K-1)}{1-q_0-q_1} \int_0^{\frac{1-q_0-q_1}{K-1}}x \left[1-\frac{(K-1)x}{1-q_0-q_1}\right]^{K-1}~{\rm d}x\\
&=\frac{1-q_0-q_1}{K^2-1}.
\end{aligned}
\end{equation}
Hence the probability that $\tilde{t}$ falls into $[{c_1}/{K^2}, {c_2}/{K^2}]$ is
\begin{align}\label{t_fallinto}
&{\rm Pr}({c_1}/{K^2}\leq \tilde{t}\leq {c_2}/{K^2})= F_{\tilde{t}}(c_2/K^2)-F_{\tilde{t}}(c_1/K^2)\notag\\
=&\!\!\left[1-\frac{c_1(K-1)}{(1-q_0-q_1)K^2}\right]^K\!\!\!\!\!\!-\left[1-\frac{c_2(K-1)}{(1-q_0-q_1)K^2}\right]^K\!\!\!\!\!\!.
\end{align}
When $K$ is large, $(K-1)/{K}\approx 1$ and $(1-1/(c'K))^{c'K}\approx 1/e(c'>0)$. Equation (\ref{t_fallinto}) can be approximated as
\begin{align}\label{t_fallinto_2}
{\rm Pr}({c_1}/{K^2}\leq \tilde{t}\leq {c_2}/{K^2})\approx  e^{-\frac{c_1}{1-q_0-q_1}}- e^{-\frac{c_2}{1-q_0-q_1}}.
\end{align}
Provided that $q_0=q_1=0$, when $c_1 = 0.1$ and $c_2 = 10$, ${\rm Pr}({0.1}/{K^2}\leq \tilde{t}\leq {10}/{K^2})\approx 0.94$; when $c_1 = 0.01$ and $c_2 = 100$, ${\rm Pr}({0.01}/{K^2}\leq \tilde{t}\leq {100}/{K^2})\approx 0.99$. It can be seen that $\tilde{t}$ falls near the order of magnitude of $K^{-2}$ with high probabilities. Thus it is reasonable that $\tilde{t}=O( K^{-2})$.

According to the definition of $p_i$ (\ref{y_probability}), equations (\ref{def_tl}) and (\ref{conjecture_t}), $c_t \propto 1-q_0-q_1$. From (\ref{simbound}), the number of quantizers $N_{\rm req}$ required for permutation matrix recovery probability is
\begin{align}\label{Kneeded}
N_{\rm req}\propto 1/{(1-q_0-q_1)^2}.
\end{align}
From (\ref{Kneeded}), one can conclude that the number of quantizers for permutation matrix recovery with high probability is ${1}/{(1-q_0-q_1)^2}$ times that of unflipped case where $q_0=q_1=0$.

\section{numerical Simulations}\label{simulation}
In this section, numerical experiments are conducted to evaluate the theoretical results. For simplicity, the distribution of noise $w_{ij}$ is selected as the Gaussian distribution $\mathcal N(0,\sigma_w^2)$.
\subsection{Parameter estimation}
For the first two experiments, we evaluate the performance of ML estimators proposed in section \ref{section_est}. Parameters are set as follows: $K=20$, $\theta=1$, $\sigma_w^2=1$, $\Delta=2$, $q_0=0.05$, $q_1=0.05$ and the tolerance parameter $\epsilon$ in Algorithm \ref{alternating_m} is $10^{-7}$. The number of Monte Carlo trials is $5000$.

\begin{figure}[h!t]
\centering
\includegraphics[width=2.8in]{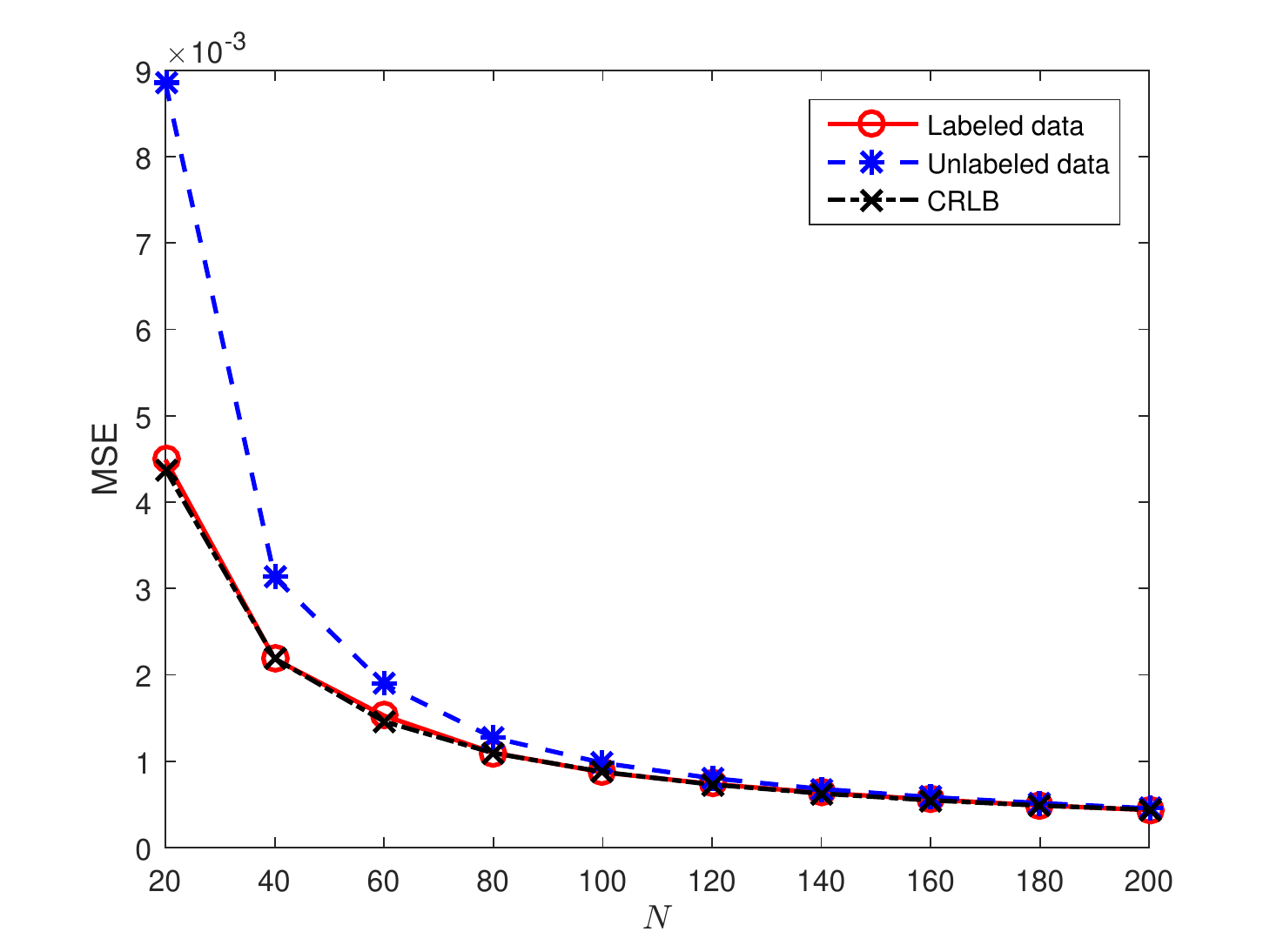}
\caption{MSE of $\theta$ vs. number of quantizers for the ML estimators from labeled and unlabeled data, compared with the CRLB (\ref{CRLBdef}) for ramp signal.}
\label{asymmetirc_h_2tau_MSE}
\end{figure}

For the first experiment, the MSE performance of Algorithm \ref{reorder} is evaluated in Fig. \ref{asymmetirc_h_2tau_MSE}. We let $\boldsymbol\tau=0.5\mathbf h$, which is a special case mentioned in Proposition \ref{simplesort}. The coefficients $\mathbf h$ is equispaced with ${\mathbf h}=[-1.50,-1.29,-1.08,\cdots,2.50]^{\rm T}$, which corresponds to a ramp signal. It can be seen that $\mathbf h$ does not satisfy the condition in Proposition \ref{statid}, thus the model may be identifiable. It can be seen that the ML estimator from labeled data always works well. Given limited number of quantizers, there is an obvious gap between the MSEs of two estimators. As the number of quantizers increases, the performance of the estimator from unlabeled data approaches that from labeled data.

\begin{figure}[h!t]
\centering
\includegraphics[width=2.8in]{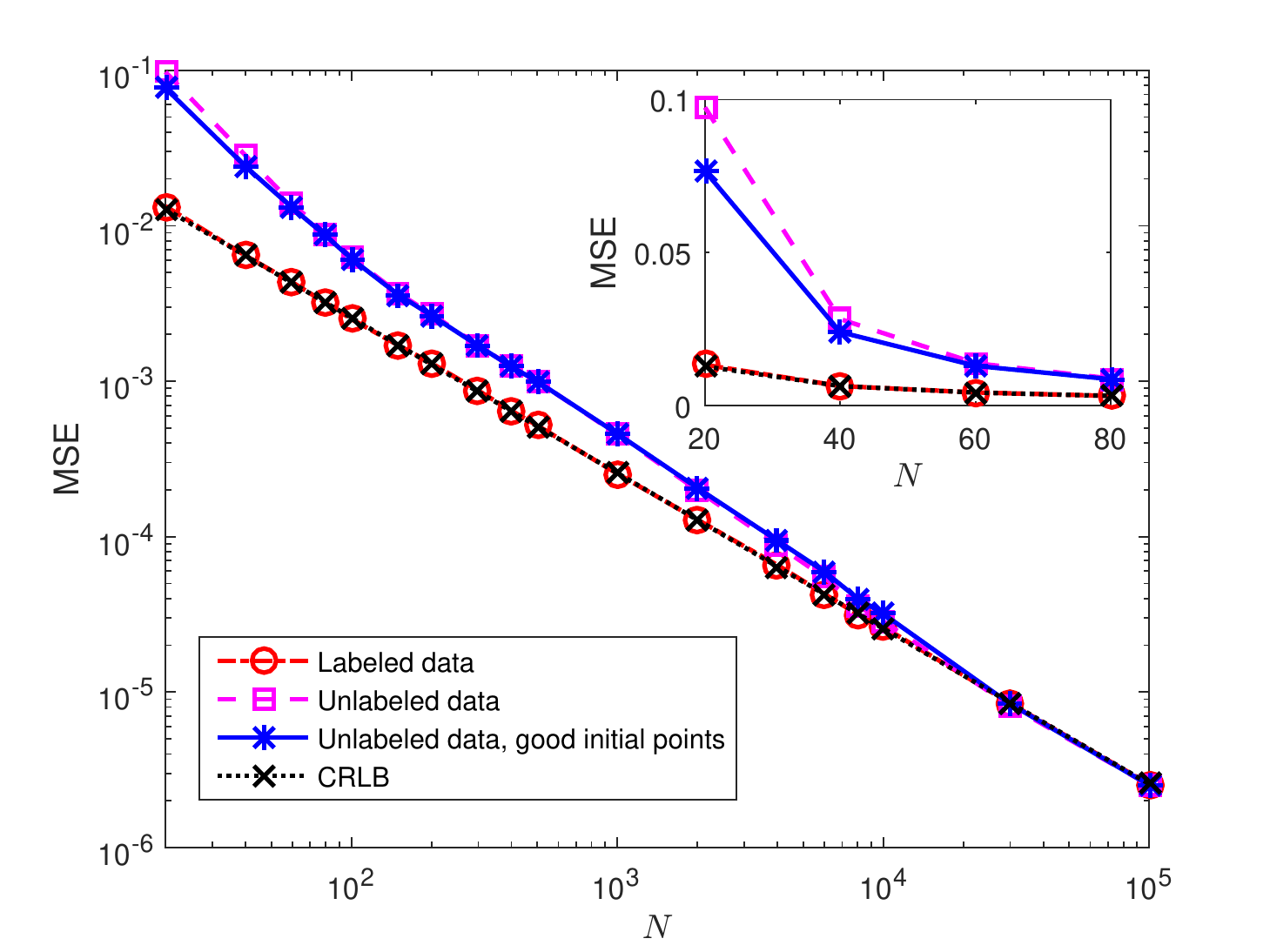}
\caption{MSE of $\theta$ vs. number of quantizers for the three ML estimators from labeled data, unlabeled data via initial points $\pm \Delta$ and unlabeled data via good initial points (\ref{goodinits}), compared with the CRLB (\ref{CRLBdef}) for sinusoidal signal.}
\label{random_h_random_tau_MSE}
\end{figure}

For the second experiment, the MSE performance of Algorithm \ref{alternating_m} (for the general case) is evaluated In Fig. \ref{random_h_random_tau_MSE}.  The elements of the vector $\mathbf h$ describe the shape of a sinusoidal signal such that $h_i=\sin (2\pi x_i)$, where $x_i$ is drawn independently and randomly from the uniform distribution $\mathcal U (0,1)$ and then sorted in ascending order. The elements of the vector ${\boldsymbol \tau}$ is drawn independently and randomly from the uniform distribution $\mathcal U (-\Delta,\Delta)$. It can be seen that when $N<80$, good initial points improve the MSE performance of the alternating maximization algorithm from unlabeled data. As $N$ increases to $80$, the MSE performances of both unlabeled ML estimators approach a common level which is larger than that achieved by the labeled data. Finally, the MSEs of both estimators from unlabeled data approach to that from labeled data around $N=3\times10^4$.


\subsection{Signal detection}

\begin{figure}[h!t]
\centering
\includegraphics[width=3.6in]{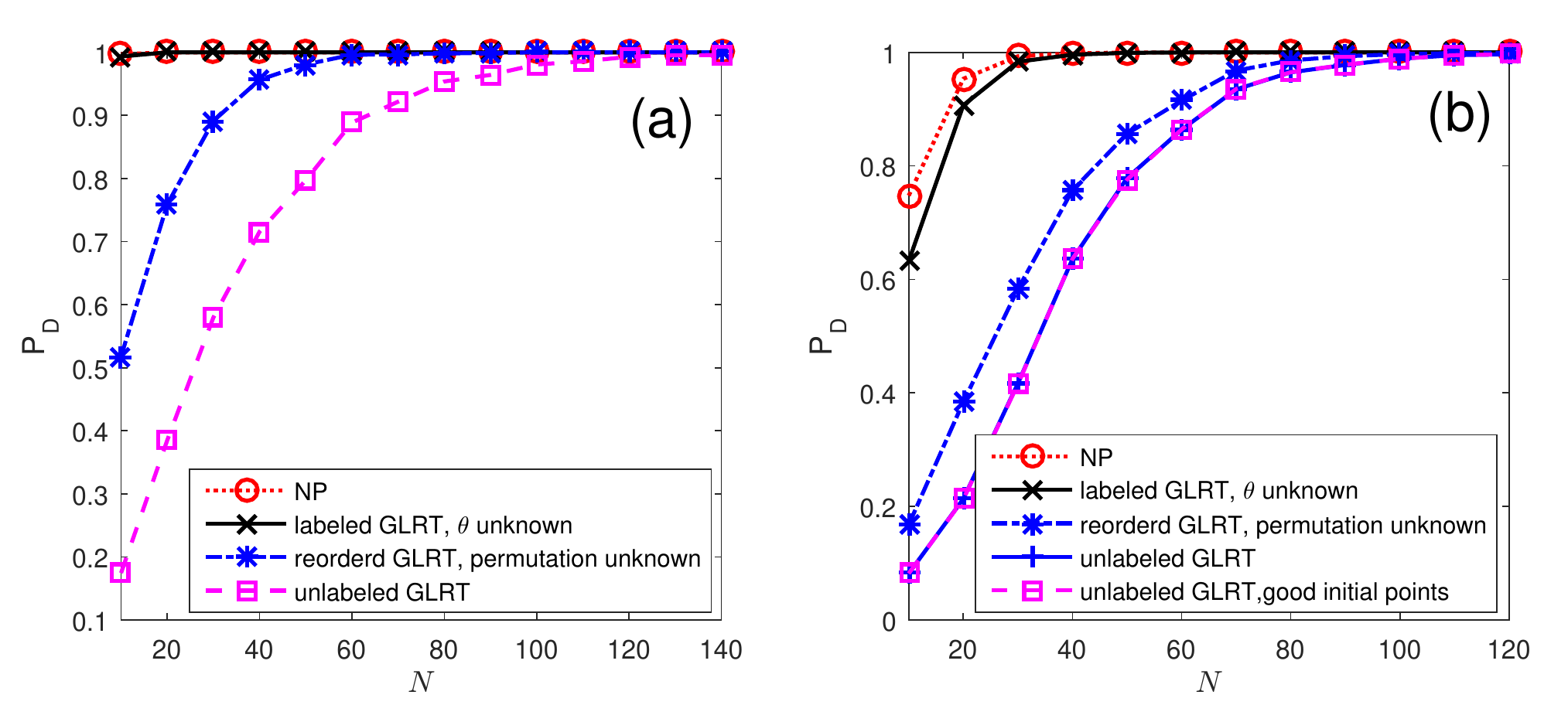}
\caption{$P_d$ vs. number of quantizers $N$ for the ramp signal in subgraph ($a$) and the sinusoidal signal in subgraph ($b$).}
\label{combine_fig7to8}
\end{figure}

In Fig. \ref{combine_fig7to8}, the relationship between $P_D$ and the number of quantizers $N$ is employed. Parameters are consistent with the first experiment, except that $\sigma_w^2=9$ and $P_{FA} = 0.05$.

In subgraph ($a$), $\mathbf h$ and $\boldsymbol\tau$  are the same as those in the first experiment. It can be seen that the number of quantizers has a significant effect on the detection probability. As $N$ increases, the performance of all the detectors improves, and the detection performance of the unlabeled GLRT approaches to that of labeled GLRT. In subgraph ($b$), $\mathbf h$ and $\boldsymbol\tau$ are the same as those in the second experiment, and similar phenomena are observed. It seems that in this case little is gained by good
initialization.

\subsection{Permutation matrix recovery}
In this subsection, the approximations for permutation matrix recovery are verified. Parameters are set as follows: $K = 20$ , $\theta = 1.5$, $\Delta = 2$, $q_0 = 0$, $q_1 = 0$ and $\sigma_w^2=1$. The number of Monte Carlo trials is $1000$.

\begin{figure}[h!t]
\centering
\includegraphics[width=2.8in]{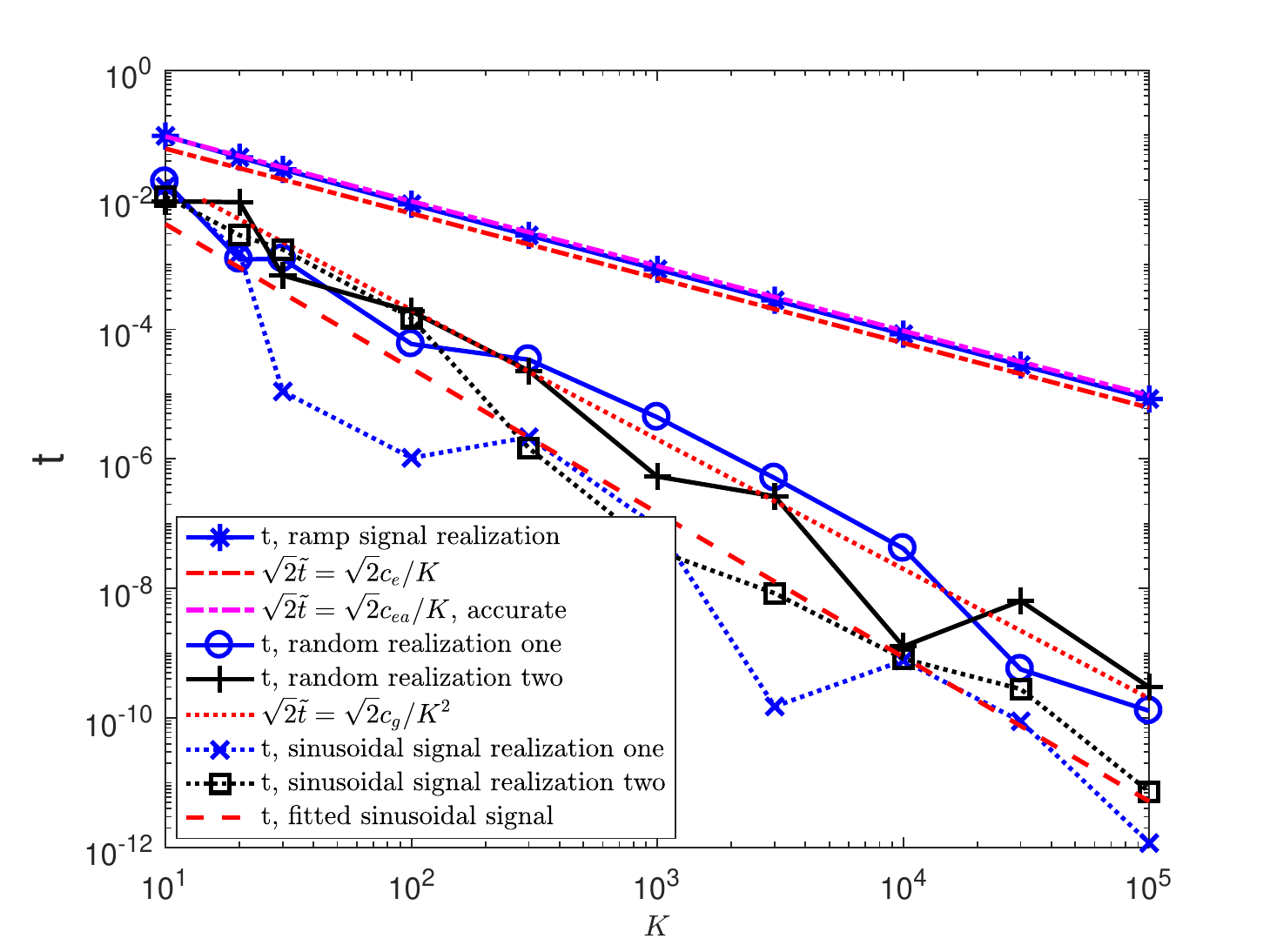}
\caption{The relationship of $t$ and $K$, including equispaced, randomly generated and sinusoidal $\mathbf h$ cases.}
\label{t_vs_N}
\end{figure}

First, the relationship of $t$ and $\tilde t$ (\ref{def_tl}) and the conjecture of $\tilde t$ (\ref{conjecture_t}) are illustrated in three cases. From Fig. \ref{t_vs_N}, one obtains that $t$ can be approximated as $\sqrt{2}{\tilde t}$ in practice. For a ramp signal, $\mathbf h = [-0.800,-0.705,-0.610,\cdots,1.000]^{\rm T}$ and $\boldsymbol\tau = 0.5 \mathbf h$. $t\approx \sqrt{2}c_e/K$ where $c_t=c_e=0.4355$ is evaluated via (\ref{c_t}). Because of the gap between $t$ and $\sqrt{2}c_e/K$, we use linear regression to fit $t$ and obtain $c_{ea}=0.6717$, which is much more accurate than $c_e$ and will be utilized later to predict the number of quantizers for permutation matrix recovery. For random generated $\mathbf h$, $\mathbf h$ is drawn from standard normal distribution and $\boldsymbol\tau = 0.5 \mathbf h$. It can be seen that $t$ can be approximated by $1/K^2$. For a sinusoidal signal, $\mathbf h$ and $\boldsymbol\tau$ are drawn in the same way of the second experiment. We use linear regression and obtain $t\approx0.71/K^{2.23}\approx {\sqrt{2}\tilde t}=\sqrt{2}c_{t,s}/K^{\alpha_{t,s}}$, $c_{t,s}=0.5020$ and $\alpha_{t,s}=2.23$.

\begin{figure}[h!t]
\centering
\includegraphics[width=3.6in]{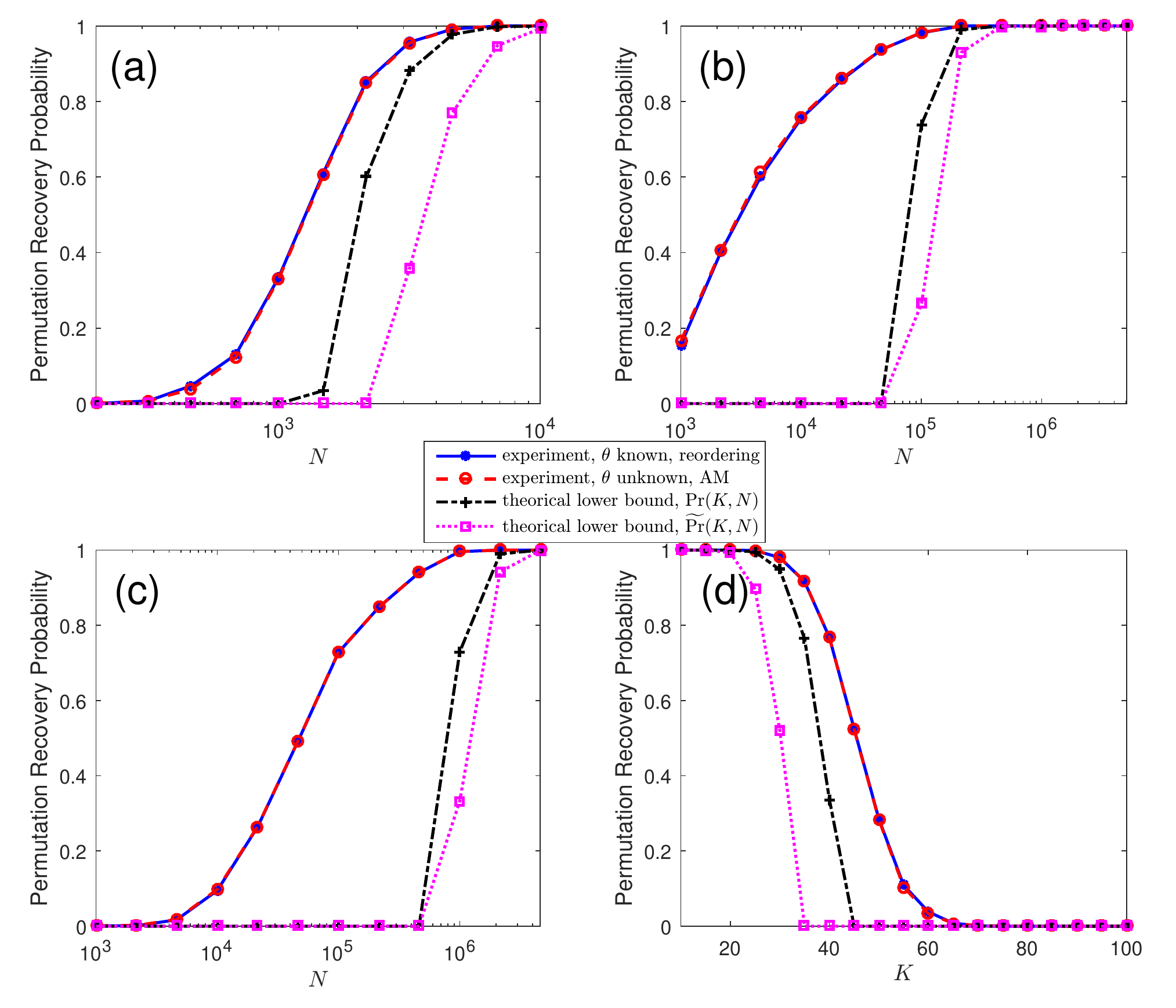}
\caption{${\rm Pr}(\hat{\boldsymbol \Pi}_{\rm ML}={\boldsymbol \Pi}_0)$ vs. $N$ or $K$ for the ramp signal in subgraph ($a$)($d$), random generated $\mathbf h$ in subgraph ($b$) and the sinusoidal signal in subgraph ($c$). ${\rm Pr}(K,N)$ and $\widetilde{\rm Pr}(K,N)$ are evaluated via (\ref{theory_approx}) and (\ref{theory_NK1}), respectively.}
\label{combine_fig10to14}
\end{figure}

Next, the empirical permutation recovery probability ${\rm Pr}(\hat{\boldsymbol \Pi}_{\rm ML}={\boldsymbol \Pi}_0)$ versus $N$ or $K$ are presented in Fig. \ref{combine_fig10to14}, and the theoretical approximations ${\rm Pr}(K,N)$ (\ref{theory_approx}) and $\widetilde{\rm Pr}(K,N)$ (\ref{theory_NK1}) are plotted for comparison. In subgraph ($a$), ($b$) and ($c$), we set $K=20$. While in subgraph ($d$), we set $N=10^4$. All $\mathbf h$ are drawn in the same way as the second experiment. We also evaluate the empirical permutation matrix recovery probability in the case of unknown $\theta$, which has negligible difference compared to that in the known $\theta$ case.

\begin{figure}[htp]
\centering
\includegraphics[width=2.8in]{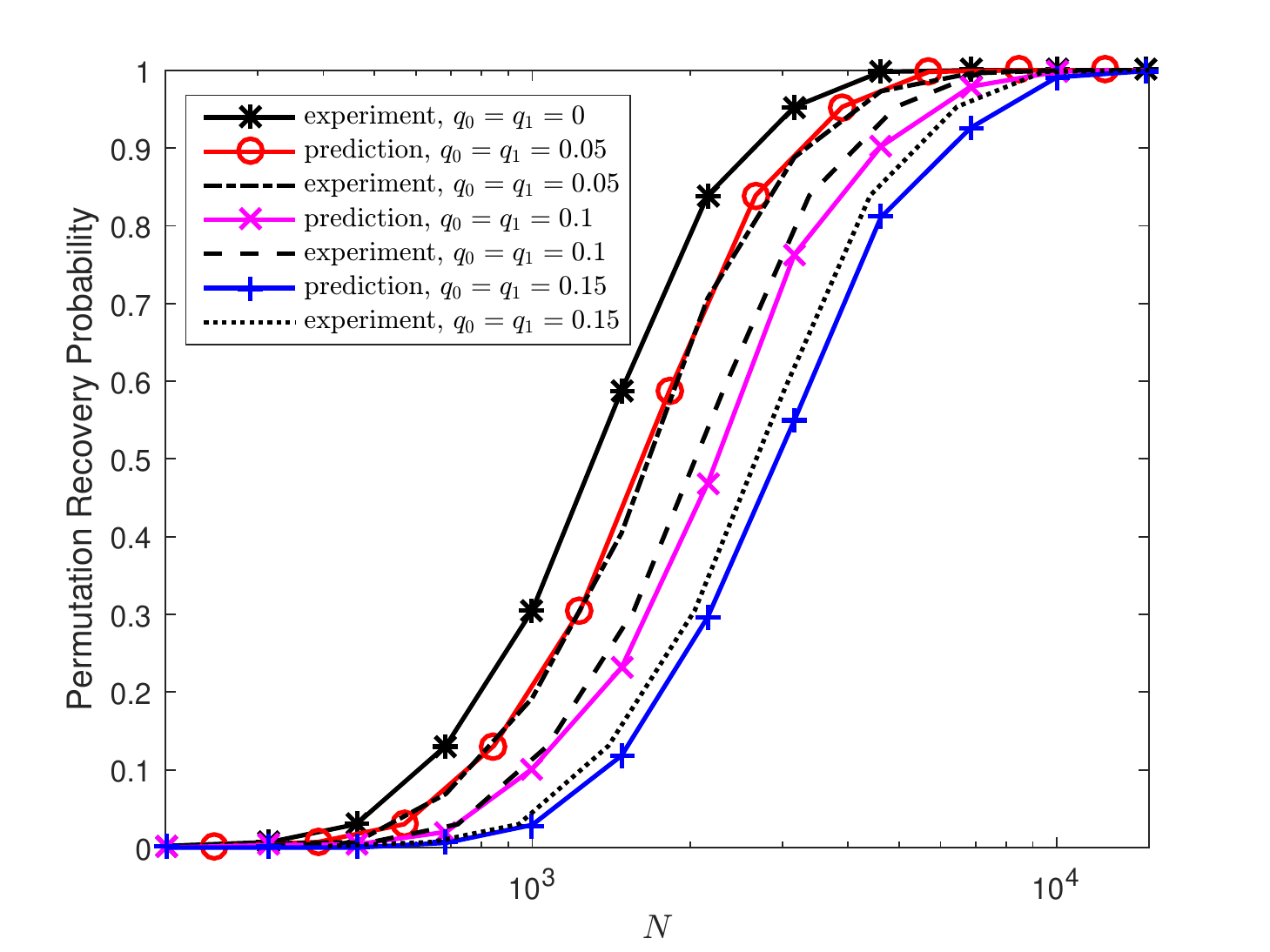}
\caption{${\rm Pr}(\hat{\boldsymbol \Pi}_{\rm ML}={\boldsymbol \Pi}_0)$ vs. number of quantizers $N$ for ramp signal under different flipping probabilities $(q_0,q_1)$.}
\label{Prec_N_q01}
\end{figure}

In subgraph($a$), it can be seen that the permutation matrix of the ramp signal can be recovered with high probability given $N\geq 5000$. From $N>\frac{1+\alpha}{c_t^2}K^{2\alpha}\ln K$ (\ref{simbound}) where $c_t=c_e=0.4355$ and $\alpha=1$, one can conclude that $N>\frac{2}{0.4355^2}K^2 \ln K|_{K=20}\approx 12636$, which is more than twice of $5000$. Utilizing the fitted parameter $c_{ea}$, one obtain a more accurate result that $N>\frac{2}{0.6717^2}K^2 \ln K|_{K=20}\approx 5312$ ensures permutation matrix recovery with high probability. For random $\mathbf h$, $N>3K^4 \ln K|_{K=20}\approx 1.438\times10^6$ ensures recovery with high probability, which is not accurate enough, as subgraph($b$) shows that $N\approx 10^5$ is enough for recovery of permutation matrix. In subgraph($c$), it is shown that $N\approx 10^6$ is enough for recovery of permutation matrix, which is also inaccurate compared to the fitted results of the sinusoidal signal $N>\frac{3.23}{0.5020^2}K^{4.46} \ln K|_{K=20}\approx 2.437\times10^7$. The numerical results show that the theoretical bound ${\rm Pr}(K,N)$ is accurate in predicting $N$ with high probability in permutation matrix recovery, which demonstrates that $\widetilde{\rm Pr}(K,N)$ may be too conservative in predicting the number of quantizers ensuring perfect permutation matrix recovery. In subgraph($d$), $10000=N>\frac{2}{0.6717^2} K^2 \ln K|_{K=26}\approx 9763$, thus $K\leq 26$ will ensure permutation matrix recovery with high probability, which is consistent with the numerical results.
In Fig. \ref{Prec_N_q01}, the relationship of flipping probabilities $(q_0, q_1)$  and number of quantizers $N_{\rm req}$ (\ref{Kneeded}) required for permutation matrix recovery with high probability is verified. Parameters are the same as those in Fig. \ref{combine_fig10to14}-($a$) except for $(q_0,q_1)$. We use the result of the experiment in which $q_0=q_1=0$ to predict those in which $q_0=q_1=0.05$, $q_0=q_1=0.1$ and $q_0=q_1=0.15$, and plot the experimental results for comparison. It can be seen that the predictions are basically consistent with the experimental results, which verifies (\ref{Kneeded}).

\section{conclusion}\label{con}
We study a scale parameter estimation and signal detection
problem from unlabeled quantized data for a canonical (known signal shape) sensing model. A sufficient condition under which the signal amplitude estimation problem can be solved efficiently is provided. It is also shown that in some settings the model can even be unidentifiable. Given that the number of quantizers is limited, the performance of the unlabeled estimator via reordering and alternating maximization algorithms is good, although there is a gap between the performances of labeled and unlabeled ML estimators. In addition, good initial points are provided to improve the performance of an alternating maximization algorithm for general estimation problems. As the number of quantizers increases, the performance of the unlabeled estimator approaches that of the labeled estimator due to the recovery of permutation matrix.

Furthermore, the performance of GLRT detector under unlabeled samples is evaluated, and numerical results show that the performance degradation of the GLRT detector under unlabeled samples is significant in noisy environments, compared to the GLRT detector with labeled samples given that the number of quantizers is small. As the number of quantizers increases, the performance of the GLRT under unlabeled samples approaches that of the GLRT detector under labeled samples. The explicit approximated permutation matrix recovery probability predicts that in order to find the true label of $K$ time indexes, the number of quantizers $N$ should be on the order of $K^{2\alpha}\log K$, where $\alpha$ is a constant depending on the signal shape and the distribution of noise.


\end{document}